\documentclass[11pt,a4paper]{article}
\usepackage[utf8]{inputenc}
\usepackage[T1]{fontenc}
\usepackage{amsmath,amssymb,amsthm,mathrsfs,bm}
\usepackage{geometry}
\usepackage{graphicx}
\usepackage{booktabs}
\usepackage{pgfplots}\pgfplotsset{compat=1.16}
\usetikzlibrary{decorations.pathreplacing}
\usepackage{hyperref}\hypersetup{colorlinks=true,linkcolor=blue,citecolor=blue,urlcolor=blue}

\newtheorem{proposition}{Proposition}

\newtheorem{remark}{Remark}

\newcommand{\R}{\mathcal{R}}

\newcommand{\Qs}{\mathcal{Q}}
\newcommand{\su}{\mathfrak{su}(1,1)}
\newcommand{\halg}{\mathfrak{h}_1}
\newcommand{\ralg}{\mathfrak{r}_1}
\newcommand{\jalg}{\mathfrak{j}_1}
\newcommand{\awalg}{\mathfrak{aw}_1}
\newcommand{\bialg}{\mathfrak{bi}_1}
\newcommand{\ket}[1]{|#1\rangle}
\newcommand{\bra}[1]{\langle#1|}
\newcommand{\ii}{\mathrm{i}}

\newcommand{\half}{\tfrac12}

\title{\bf The supersymmetric Scarf~I Hamiltonian with reflections and its discrete para--Bannai--Ito version}
\author{
St\'ephane Z. Beaulac \textsuperscript{1}\thanks{E-mail: stephane.jr.beaulac@umontreal.ca},\quad Nicolas Cramp\'e\textsuperscript{1,2}\thanks{E-mail: crampe1977@gmail.com},\quad Quentin Labriet\textsuperscript{1}\thanks{E-mail: quentin.labriet@umontreal.ca},\\[3pt]
Lucia Morey\textsuperscript{1}\thanks{E-mail: lucia.morey@umontreal.ca},\quad Marlon Josue Rivera Valladares\textsuperscript{1}\thanks{E-mail: marlon.josue.rivera.valladares@umontreal.ca},\quad Luc Vinet\textsuperscript{1}\thanks{E-mail: luc.vinet@umontreal.ca}\\[6pt]
\small\textsuperscript{1}Centre de recherches math\'ematiques, Universit\'e de Montr\'eal, P.O.\ Box 6128,\\
\small Centre-ville Station, Montr\'eal (Qu\'ebec), H3C 3J7, Canada\\[2pt]
\small\textsuperscript{2}Laboratoire d'Annecy de Physique Th\'eorique, 9 Chemin de Bellevue, BP 110,\\
\small Annecy-le-Vieux, F-74941 Annecy Cedex, France
}
\date{\today}

\begin{document}
\maketitle

\begin{abstract}
The para--Bannai--Ito polynomials are orthogonal on a Bannai--Ito bi--lattice that is
\emph{two--sided} --- it extends on both sides of a central value $c_0$ --- and depend, besides
the two grid parameters, on an isospectral deformation parameter.
Their Jacobi matrix $J$ is
invariant under the reversal of the sites at one value of that parameter only, the
persymmetric point, and we determine its continuum limit there. Because the grid is
two--sided, the natural object is the shifted operator $Q=J-c_0$, and the limit is taken about
the centre of the band: $Q$ contracts to a first--order Dunkl operator, a supercharge, and
$H=Q^2$ to the generalized P\"oschl--Teller (Scarf~I) Hamiltonian, with little $-1$ Jacobi
eigenfunctions --- supersymmetric quantum mechanics with reflections, the reflection entering
the square root of the Hamiltonian. The second grid parameter adds a reflection term to the
Hamiltonian itself, and the bi--lattice is, up to a shift, the spectrum of the limiting
supercharge. The convergence of the eigenvectors is proved directly on the polynomials. On this
grid the persymmetric model transfers states perfectly, under an arithmetic condition on the
parameters, but never revives fractionally, revival being the work of the deformation; both
properties pass to the limit. These results are for an even number of sites; for an odd number
of sites the limit is instead the matrix supersymmetric pair of P\"oschl--Teller Hamiltonians,
with the reflection in the Hamiltonian as a Dunkl potential. The bispectral pair of the discrete
model goes over to the Dunkl realization of the anticommutator algebra of the little $-1$ Jacobi
polynomials.
\end{abstract}

\tableofcontents

\section{Introduction}
\label{sec:intro}

Restricted to one excitation, the Hamiltonian of an $XX$ spin chain is a real symmetric
tridiagonal matrix: the local magnetic fields on the diagonal, the couplings between
neighbouring sites next to it. When these entries are the recurrence coefficients $B_n$ and
$\sqrt{U_n}$ of a family of orthonormal polynomials,
$x\,p_n=\sqrt{U_{n+1}}\,p_{n+1}+B_n\,p_n+\sqrt{U_n}\,p_{n-1}$, the matrix is the Jacobi
matrix $J$ of the family: its eigenvalues are the orthogonality grid and its eigenvectors the
polynomials evaluated on the grid. The transport of an excitation is then read off these
spectral data, and families for which an excitation launched at one end is transferred
perfectly to the other, or revived fractionally at the two ends, have been identified and constructed \cite{VZ,paraRacahChain,GVZ}. The para--Krawtchouk \cite{VZ}, para--Racah \cite{paraRacah}
and para--Bannai--Ito \cite{PBI} polynomials,%
\footnote{The prefix, taken over from the para--Krawtchouk polynomials, where it records that
the spectrum is that of the parabose oscillator \cite{Rosenblum}, is unrelated to the
para--orthogonal polynomials on the unit circle of \cite{JNT}, the combinations
$\Phi_n(z)+\tau\Phi_n^{*}(z)$, $|\tau|=1$, used in Szeg\H{o} quadrature.}
three families orthogonal on bi--lattices whose common structure is studied in
\cite{BCLMNV}, are among them. Three companion papers, of which this is the last, determine
the continuum limits of their Hamiltonians as the number of sites grows. The first two limits
are established in \cite{PKcont,PRcont}: the para--Krawtchouk Hamiltonian, whose spectrum is
a \emph{linear} bi--lattice, contracts to the singular (isotonic) oscillator on the line, and
the para--Racah Hamiltonian, whose spectrum is a \emph{quadratic} bi--lattice, to the
trigonometric P\"oschl--Teller Hamiltonian on a bounded interval. In both cases the limit is a
shape--invariant Schr\"odinger operator, obtained through the classical confluences
Hahn~$\to$~Laguerre and Racah~$\to$~Jacobi, and the mirror symmetry of the discrete model
passes to a reflection that commutes with the limiting Hamiltonian.

This paper treats the third, the para--Bannai--Ito Hamiltonian, a matrix acting on
$N+1$ sites, $N=2j+1$ with $j$ even in the main text (the other parity classes are treated in
\S\ref{sec:even}). The para--Bannai--Ito polynomials \cite{PBI} arise as the $q\to-1$ limit of the
$q$--para--Racah polynomials. They are orthogonal on a linear bi--lattice, the union of two
\emph{sublattices}, which is \emph{two--sided}: its points lie on both sides of a central value
$c_0$, each sublattice having an arm on each side, each arm an arithmetic progression of unit
step, and on each side the two arms interlace. The grid depends on
two parameters, $a$ and $b$: $a$ sets the gap between the two sides, and $b$ the spacing within
each side, where the gaps between consecutive points alternate between $\tfrac{1-b}2$ and
$\tfrac{1+b}2$ (\S\ref{sec:chain}). A third parameter $\alpha\in(0,1)$ leaves the grid
unchanged and deforms the Jacobi matrix, which is persymmetric --- invariant under the reversal
of the sites --- exactly at $\alpha=\half$, the \emph{persymmetric point}, where we shall
mostly work. Like their $q=-1$ ancestors the Bannai--Ito polynomials \cite{BI,TVZ,DGTVZ}, the
para--Bannai--Ito polynomials are eigenfunctions \cite{PBI} of a first--order \emph{Dunkl} operator
\cite{Dunkl}, i.e.\ a differential--difference operator involving the reflection $\R$. We
answer the same question as before: \emph{what is the continuum limit of the para--Bannai--Ito
Hamiltonian $J$?}

The answer is a \emph{supersymmetric system with reflections} in the sense of \cite{PVZ},
and three features set it apart from the first two limits, all three stemming from the
geometry of the $q=-1$ grid.

First, since the grid straddles its centre $c_0$, with $x_{\min}+x_{\max}=2c_0$, the shifted
operator $Q=J-c_0$ has eigenvalues on both sides of zero, so $Q$ is not itself a Schr\"odinger
operator but a \emph{square root} of one: the natural Hamiltonian is $H=Q^2$. This is the
finite--dimensional counterpart of the ``square root of the Schr\"odinger operator'' attached
to the little $-1$ Jacobi polynomials in \cite[\S7]{VZm1}.

Second, and consequently, the contraction is taken about the \emph{centre} of the band rather
than a band edge. At the persymmetric point and for $b=0$ the shifted matrix $Q$ has no diagonal
at all: it is chiral, its spectrum is symmetric about zero, and the states of small $|Q|$ sit in
the middle of the band, where the dispersion of a tridiagonal matrix crosses the energy
\emph{linearly} and not quadratically as at an edge --- a \emph{Dirac point}, in the language of
band theory, explained in \S\ref{sec:eqlimit}. There the recurrence is best read not as a
discrete Schr\"odinger equation but as a discrete \emph{Dirac} equation for a two--component
amplitude, and $Q$ contracts to a \emph{first--order} operator. It is $H=Q^2$ that is second
order.

Third, the limiting first--order operator is of \emph{Dunkl type}: it contains the reflection $\R$.
At the persymmetric point $J$ commutes with the reversal of the sites, so that its eigenvectors
are even or odd under it, and on a bi--lattice the parity is the same for all the eigenvalues
of a sublattice; the two
eigenspaces of the site reversal, which we call the two \emph{channels}, thus carry the two
sublattices of the spectrum, and in the limit they become two families of eigenfunctions of
definite parity. In the para--Krawtchouk limit the reflection enters only through the choice of
self--adjoint extension at the origin, and in the para--Racah limit as the $\mathbb{Z}_2$
grading of the two channels; here it enters the dynamics directly --- as the supercharge, the
square root of the Hamiltonian. The limit is the extended Scarf~I Hamiltonian with a Dunkl
supercharge of \cite{PVZ}, its eigenfunctions the little $-1$ Jacobi functions, and the mirror
symmetry combines the two channels into a single supersymmetric system.

The computation follows the pattern of the two companions. In \S\ref{sec:chain} we define the
para--Bannai--Ito Hamiltonian from the recurrence coefficients of the polynomials of \cite{PBI}
and describe its two--sided grid. In \S\ref{sec:eqlimit} we recast the recurrence as
a discrete Dirac system and contract it to the Dunkl supercharge and its square. In \S\ref{sec:polylimit} we take the limit directly on the polynomials and prove that the
eigenvectors converge to the little $-1$ Jacobi functions; these are the eigenfunctions of
the limiting supercharge, whose spectrum is, up to a shift, the bi--lattice itself.
Section~\ref{sec:reflection} treats the supersymmetry with reflections and the two channels,
including the second grid parameter, which adds a reflection term to the Hamiltonian, and
\S\ref{sec:even} treats the three other parity classes of the para--Bannai--Ito polynomials:
the parity of $j$ is immaterial, but for even $N$ (an odd number of sites) the Hamiltonian
contracts to a different realization of the same supersymmetry, with the same eigenvalues ---
the ordinary matrix one, with the reflection in the Hamiltonian instead of the supercharge.
Section~\ref{sec:algebra} determines what the two bispectral operators of the discrete model
become in the limit, namely the Dunkl realization of the anticommutator algebra of the little
$-1$ Jacobi polynomials; \S\ref{sec:compare} places the result in the trilogy; and
\S\ref{sec:FR} determines when the discrete model transfers or revives and what of this
survives the limit. Appendix~\ref{app:confluence} supplies the hypergeometric identities ---
Whipple's transformation, the profile of the polynomials, the connection and contiguous
relations --- on which the convergence proof of \S\ref{sec:polylimit} rests.

\section{The para--Bannai--Ito polynomials}
\label{sec:chain}

Let $N=2j+1$ with $j$ even (the three other parity classes are treated in \S\ref{sec:even}).
The monic para--Bannai--Ito polynomials $P_n(x)$ are defined \cite{PBI} by the three--term
recurrence relation
\begin{equation}
\label{eq:rec}
xP_n(x)=P_{n+1}(x)+B_nP_n(x)+U_nP_{n-1}(x),\qquad
B_n=\frac{a+b-j-1}{4}-A_n-C_n,\quad U_n=A_{n-1}C_n,
\end{equation}
with $P_{-1}(x)=0$, $P_0(x)=1$ and the parity--dependent coefficients
\begin{equation}
\label{eq:AC}
\begin{aligned}
A_n&=\begin{cases}
\tfrac14(n-j+a), & n\text{ even},\ n\neq j,\\[2pt]
\tfrac14\dfrac{(n-2j-1)(n-j+b)}{n-j}, & n\text{ odd},\\[6pt]
\tfrac12\,\alpha\,a, & n=j,
\end{cases}
\\[4pt]
C_n&=\begin{cases}
-\tfrac14\dfrac{n(n-j-1-b)}{n-j-1}, & n\text{ even},\\[6pt]
-\tfrac14(n-j-1-a), & n\text{ odd},\ n\neq j+1,\\[2pt]
\tfrac12(1-\alpha)a, & n=j+1.
\end{cases}
\end{aligned}
\end{equation}
Here $a,b$ are grid parameters and $\alpha\in(0,1)$ is an ``isospectral'' parameter, in a sense
made precise in \S\ref{sec:FR}; the recurrence defines a genuine real Jacobi matrix, with
$U_n>0$ for $n=1,\dots,N$, when \cite{PBI}
\begin{equation}
\label{eq:pos}
|a|\ge j+1,\qquad |b|<1,\qquad 0<\alpha<1
\end{equation}
(at $|b|=1$ or $\alpha\in\{0,1\}$ one of the $U_n$ vanishes and the matrix decouples into two
blocks; we exclude these endpoints). We take $a\le-(j+1)$ and write $a=-(j+1)-2\mu$ with
$\mu\ge0$ fixed; replacing $a$ by $-a$ exchanges the two sublattices of the grid \eqref{eq:grid}
below, so nothing is lost. The bound \eqref{eq:pos} forces $a$ to scale with the number of
sites, in contrast to the fixed parameters of the para--Krawtchouk and para--Racah Hamiltonians.
The orthonormal polynomials are $p_n=P_n/h_n$ with $h_n=\sqrt{U_1\cdots U_n}$ ($h_0=1$), and
\eqref{eq:rec} becomes
\begin{equation}
\label{eq:recnorm}
xp_n(x)=\sqrt{U_{n+1}}\,p_{n+1}(x)+B_n\,p_n(x)+\sqrt{U_n}\,p_{n-1}(x).
\end{equation}
Let $\ket{e_n}$, $n=0,\dots,N$, be the canonical basis of $\mathbb C^{N+1}$ --- the
\emph{sites} --- and $J$ the Jacobi matrix of the family, the real symmetric tridiagonal
matrix with
\begin{align}
&\langle e_n|J|e_n\rangle=B_n\,,\\
&\langle e_{n}|J|e_{n+1}\rangle=\langle e_{n+1}|J|e_{n}\rangle=\sqrt{U_{n+1}}\,.
\end{align}
This is the para--Bannai--Ito Hamiltonian. Its eigenvalues are the zeros $x_s$ of $P_{N+1}$,
and by \eqref{eq:recnorm} the normalized eigenvector with $x_s$ as eigenvalue is
\begin{equation}
\label{eq:eigvec}
\ket{s}=\sum_{n=0}^N\sqrt{w_s}\,p_n(x_s)\,\ket{e_n},\qquad J\ket{s}=x_s\ket{s},
\end{equation}
where the $w_s>0$ are the weights of the discrete orthogonality relation
\begin{equation}
\sum_{s=0}^{N}w_s\,p_n(x_s)p_{n'}(x_s)=\delta_{nn'}\,,
\end{equation}
normalized by $\sum_sw_s=1$.

\paragraph{The two--sided grid.}
The eigenvalues form the \emph{two--sided} bi--lattice \cite{PBI}
\begin{equation}
\label{eq:grid}
x_{2s}=-(-1)^s\,\frac{2s-j-b+a}{4}-\tfrac14,\qquad
x_{2s+1}=-(-1)^s\,\frac{2s-j-b-a}{4}-\tfrac14,\qquad s=0,\dots,j,
\end{equation}
two interlaced linear lattices, the \emph{sublattices} $\{x_{2s}\}$ (even) and $\{x_{2s+1}\}$
(odd), each running to the left and to the right of a common centre (the union of four
arithmetic progressions, a ``linear quadri--lattice'' in the terminology of \cite{PBI}). Note
that \eqref{eq:grid} labels the points by sublattice, not by size. The grid straddles the point
\begin{equation}
\label{eq:center}
c_0=\frac{b-1}{4}:
\end{equation}
a direct computation gives $x_{\min}+x_{\max}=2c_0$, so $x_{\min}<c_0<x_{\max}$ and the grid
spreads on \emph{both} sides of $c_0$; at $b=0$ it is moreover \emph{fully} symmetric about
$c_0$, invariant under $x\mapsto 2c_0-x$. (The grid is the same for all $\alpha$: this
parameter acts on the eigenvectors, not on the spectrum, see \S\ref{sec:FR}.) The two grid
parameters have distinct geometrical roles \cite[App.~C]{PBI}. With $a=-(j+1)-2\mu$, and in
terms of the signed distance $q=x-c_0$ to the centre, \eqref{eq:grid} reads
\begin{equation}
\label{eq:gridq}
\begin{aligned}
x_{2s}-c_0&=(-1)^{\ell}\,\frac{2\ell+1+2\mu+b}{4}-\frac b4&&(\ell=j-s),\\
x_{2s+1}-c_0&=-(-1)^{\ell}\,\frac{2\ell+1+2\mu-b}{4}-\frac b4&&(\ell=s),
\end{aligned}
\end{equation}
so that the points to the right of the centre are $\tfrac{1+2\mu}{4}+k$ (even sublattice,
$\ell=2k$) and $\tfrac{3+2\mu-2b}{4}+k$ (odd sublattice, $\ell=2k+1$), $k=0,1,\dots$, and those
to the left are $-\tfrac{1+2\mu}{4}-k$ (odd sublattice, $\ell=2k$) and $-\tfrac{3+2\mu+2b}{4}-k$
(even sublattice, $\ell=2k+1$) --- the images of the former under $q\mapsto-q$ combined with
$b\mapsto-b$. Each side is thus the union of two arithmetic progressions of unit step --- one
arm of each sublattice --- the second being the first shifted by $\tfrac{1-b}{2}\in(0,1)$ on
the right and by $\tfrac{1+b}{2}$ on the left, so that the two progressions interlace
(Figure~\ref{fig:grid}). Thus $2\mu$ (equivalently $a$) sets the
\emph{gap} $\tfrac{1+2\mu}{2}$ between the two innermost points, across the centre, while $b$
sets the \emph{spacing} within each side, where the gaps between consecutive points alternate
between $\tfrac{1-b}{2}$ and $\tfrac{1+b}{2}$: at $b=0$ the points on each side are equally
spaced by $\tfrac12$, and as $|b|\to1$, the end of the range \eqref{eq:pos}, pairs
of points coalesce. These two roles reappear in the limit as, respectively, the depth of the
potential and the amplitude of the reflection. Equivalently, in terms of $x+\tfrac14=q+\tfrac b4$,
\begin{equation}
\label{eq:towers}
x_{2s}+\tfrac14=(-1)^{\ell}\,\frac{2\ell+1+2\mu+b}{4}\ \ (\ell=j-s),\qquad
x_{2s+1}+\tfrac14=(-1)^{\ell+1}\,\frac{2\ell+1+2\mu-b}{4}\ \ (\ell=s):
\end{equation}
counted from the centre outwards, each sublattice is a \emph{tower} $\tfrac14(2\ell+1+2\mu\pm b)$,
$\ell=0,1,2,\dots$, with alternating signs, the even sublattice with $+b$ and the odd one with
$-b$. The integer $\ell$, the position of a point in its tower, will be called its \emph{level};
it is the label that survives the limit.

\begin{remark}
The points of the grid at a bounded distance from $c_0$ --- those of bounded level --- do not
depend on $j$ (Figure~\ref{fig:grid}(c)): increasing $N$ adds points at the two ends of the
grid and leaves the others where they are. This is what allows the limit $N\to\infty$ to be
taken eigenvalue by eigenvalue, as in the two companions: the eigenvector belonging to a given
$q=x_s-c_0$ --- a value of order one in the middle of a spectrum of width $O(N)$ --- can be
followed as $N$ grows, and it converges, in the rescaled site variable, to an eigenfunction of
the limiting operator (\S\S\ref{sec:eqlimit}--\ref{sec:polylimit}), whose spectrum is the
union of the two infinite towers.
\end{remark}

\paragraph{The site reversal and the persymmetric point.}
Let $\R_c$ denote the site reversal, $\R_c\ket{e_n}=\ket{e_{N-n}}$. From \eqref{eq:AC} one reads
$A_n=C_{N-n}$ for every $n\neq j$, whereas $A_j=\tfrac12\alpha a$ and $C_{N-j}=C_{j+1}=\tfrac12(1-\alpha)a$
coincide exactly when $\alpha=\half$. Consequently the couplings are palindromic for every
$\alpha$, $U_{N+1-n}=A_{N-n}C_{N+1-n}=C_nA_{n-1}=U_n$ (trivially so at the central bond
$n=j+1$), while the diagonal satisfies $B_{N-n}=B_n$ for $n\neq j,j+1$ and
$B_j-B_{j+1}=C_{j+1}-A_j=\tfrac12(1-2\alpha)a$. Thus $J$
is \emph{persymmetric}, $\R_c J\R_c=J$, exactly at $\alpha=\half$, which we call the
\emph{persymmetric point}; this is the symmetry that underlies state transfer (\S\ref{sec:FR}),
and we shall mostly work there. At $\alpha=\half$, $\R_c$ commutes with $J$ and, the spectrum
being simple, each eigenvector is a mirror eigenvector, $\R_c\ket{s}=\epsilon_s\ket{s}$ with
$\epsilon_s=\pm1$. To determine the signs, list the eigenvalues in increasing order,
$x_{(0)}<x_{(1)}<\dots<x_{(N)}$ (recall that \eqref{eq:grid} labels them by sublattice, not by
size). Reading the mirror relation at the first site gives $\epsilon_{(s)}=p_N(x_{(s)})$, and
$p_N$ takes alternating signs at the $N+1$ points $x_{(s)}$, which interlace its $N$ zeros,
with $p_N(x_{(N)})>0$ because all its zeros lie below $x_{(N)}$ and its leading coefficient is
positive \cite{VZ,GTVZ}; hence $\epsilon_{(s)}=(-1)^{N+s}$. On the two--sided grid
\eqref{eq:grid} the two sublattices alternate along the ordered spectrum \cite{PBI} --- this is
visible on Figure~\ref{fig:grid}(a) --- and the largest eigenvalue, $x_{2s}$ with $\ell=j-s=j$
even, belongs to the even sublattice; since $\epsilon_{(s)}$ alternates along the ordered
spectrum and equals $+1$ at the top,
\begin{equation}
\label{eq:eps}
\epsilon=+1\ \text{on the even sublattice }\{x_{2s}\},\qquad
\epsilon=-1\ \text{on the odd sublattice }\{x_{2s+1}\}:
\end{equation}
the mirror parity of an eigenvector is the sublattice of its eigenvalue. Read site by site,
$\R_c\ket{s}=\epsilon_s\ket{s}$ is $p_{N-n}(x_s)=\epsilon_s\,p_n(x_s)$: the amplitudes on the
right half of the lattice are those on the left half up to the sign $\epsilon_s$.

\begin{figure}[ht]
\centering
\begin{tikzpicture}[x=1.60cm,y=1cm,font=\scriptsize]
\draw[->,thick] (-3.700,0) -- (3.850,0) node[right] {$q$};
\draw[dashed] (0,-1.05) -- (0,1.0) node[above] {$c_0$};
\draw[blue!35,very thin] (-3.600,0.55) -- (3.600,0.55);
\draw[red!35,very thin] (-3.600,-0.55) -- (3.600,-0.55);
\draw[blue!50,very thin] (3.400,0) -- (3.400,0.55);
\fill[blue] (3.400,0.55) circle (1.7pt);
\draw[blue!50,very thin] (-3.100,0) -- (-3.100,0.55);
\fill[blue] (-3.100,0.55) circle (1.7pt);
\draw[blue!50,very thin] (2.400,0) -- (2.400,0.55);
\fill[blue] (2.400,0.55) circle (1.7pt);
\draw[blue!50,very thin] (-2.100,0) -- (-2.100,0.55);
\fill[blue] (-2.100,0.55) circle (1.7pt);
\draw[blue!50,very thin] (1.400,0) -- (1.400,0.55);
\fill[blue] (1.400,0.55) circle (1.7pt);
\draw[blue!50,very thin] (-1.100,0) -- (-1.100,0.55);
\fill[blue] (-1.100,0.55) circle (1.7pt);
\draw[blue!50,very thin] (0.400,0) -- (0.400,0.55);
\fill[blue] (0.400,0.55) circle (1.7pt);
\draw[red!50,very thin] (-0.400,0) -- (-0.400,-0.55);
\fill[red] (-0.400,-0.55) +(-1.7pt,-1.7pt) rectangle +(1.7pt,1.7pt);
\draw[red!50,very thin] (0.700,0) -- (0.700,-0.55);
\fill[red] (0.700,-0.55) +(-1.7pt,-1.7pt) rectangle +(1.7pt,1.7pt);
\draw[red!50,very thin] (-1.400,0) -- (-1.400,-0.55);
\fill[red] (-1.400,-0.55) +(-1.7pt,-1.7pt) rectangle +(1.7pt,1.7pt);
\draw[red!50,very thin] (1.700,0) -- (1.700,-0.55);
\fill[red] (1.700,-0.55) +(-1.7pt,-1.7pt) rectangle +(1.7pt,1.7pt);
\draw[red!50,very thin] (-2.400,0) -- (-2.400,-0.55);
\fill[red] (-2.400,-0.55) +(-1.7pt,-1.7pt) rectangle +(1.7pt,1.7pt);
\draw[red!50,very thin] (2.700,0) -- (2.700,-0.55);
\fill[red] (2.700,-0.55) +(-1.7pt,-1.7pt) rectangle +(1.7pt,1.7pt);
\draw[red!50,very thin] (-3.400,0) -- (-3.400,-0.55);
\fill[red] (-3.400,-0.55) +(-1.7pt,-1.7pt) rectangle +(1.7pt,1.7pt);
\draw (3.400,-0.06) -- (3.400,0.06);
\draw (-3.100,-0.06) -- (-3.100,0.06);
\draw (2.400,-0.06) -- (2.400,0.06);
\draw (-2.100,-0.06) -- (-2.100,0.06);
\draw (1.400,-0.06) -- (1.400,0.06);
\draw (-1.100,-0.06) -- (-1.100,0.06);
\draw (0.400,-0.06) -- (0.400,0.06);
\draw (-0.400,-0.06) -- (-0.400,0.06);
\draw (0.700,-0.06) -- (0.700,0.06);
\draw (-1.400,-0.06) -- (-1.400,0.06);
\draw (1.700,-0.06) -- (1.700,0.06);
\draw (-2.400,-0.06) -- (-2.400,0.06);
\draw (2.700,-0.06) -- (2.700,0.06);
\draw (-3.400,-0.06) -- (-3.400,0.06);
\node[blue,anchor=east] at (-3.750,0.55) {even sublattice};
\node[red,anchor=east] at (-3.750,-0.55) {odd sublattice};
\node[blue,anchor=south west,inner sep=1pt] at (0.400,0.08) {$\frac{1+2\mu}{4}$};
\node[blue,anchor=south east,inner sep=1pt] at (-1.100,0.08) {$-\frac{3+2\mu+2b}{4}$};
\node[red,anchor=north west,inner sep=1pt] at (0.700,-0.08) {$\frac{3+2\mu-2b}{4}$};
\node[red,anchor=north east,inner sep=1pt] at (-0.400,-0.08) {$-\frac{1+2\mu}{4}$};
\draw[decorate,decoration={brace,amplitude=3pt}] (0.400,0.78) -- (0.700,0.78) node[midway,above=2pt] {$\tfrac{1-b}{2}$};
\draw[decorate,decoration={brace,amplitude=3pt}] (0.700,0.78) -- (1.400,0.78) node[midway,above=2pt] {$\tfrac{1+b}{2}$};
\draw[decorate,decoration={brace,amplitude=3pt}] (-1.100,0.78) -- (-0.400,0.78) node[midway,above=2pt] {$\tfrac{1+b}{2}$};
\draw[decorate,decoration={brace,amplitude=3pt}] (-1.400,0.78) -- (-1.100,0.78) node[midway,above=2pt] {$\tfrac{1-b}{2}$};
\draw[decorate,decoration={brace,amplitude=3pt},blue] (2.400,0.78) -- (3.400,0.78) node[midway,above=2pt] {$1$};
\draw[decorate,decoration={brace,mirror,amplitude=3pt},red] (-3.400,-0.78) -- (-2.400,-0.78) node[midway,below=2pt] {$1$};
\draw[decorate,decoration={brace,mirror,amplitude=3pt}] (-0.400,-0.78) -- (0.400,-0.78) node[midway,below=2pt] {$\tfrac{1+2\mu}{2}$};
\node[anchor=west,font=\small] at (-5.600,1.45) {(a) $N=13$ ($j=6$), $\mu=0.3$, $b=0.4$};
\end{tikzpicture}

\vspace{2pt}

\begin{tikzpicture}[x=1.60cm,y=1cm,font=\scriptsize]
\draw[->,thick] (-3.700,0) -- (3.850,0) node[right] {$q$};
\draw[dashed] (0,-1.05) -- (0,1.0) node[above] {$c_0$};
\draw[blue!35,very thin] (-3.600,0.55) -- (3.600,0.55);
\draw[red!35,very thin] (-3.600,-0.55) -- (3.600,-0.55);
\draw[blue!50,very thin] (3.400,0) -- (3.400,0.55);
\fill[blue] (3.400,0.55) circle (1.7pt);
\draw[blue!50,very thin] (-2.900,0) -- (-2.900,0.55);
\fill[blue] (-2.900,0.55) circle (1.7pt);
\draw[blue!50,very thin] (2.400,0) -- (2.400,0.55);
\fill[blue] (2.400,0.55) circle (1.7pt);
\draw[blue!50,very thin] (-1.900,0) -- (-1.900,0.55);
\fill[blue] (-1.900,0.55) circle (1.7pt);
\draw[blue!50,very thin] (1.400,0) -- (1.400,0.55);
\fill[blue] (1.400,0.55) circle (1.7pt);
\draw[blue!50,very thin] (-0.900,0) -- (-0.900,0.55);
\fill[blue] (-0.900,0.55) circle (1.7pt);
\draw[blue!50,very thin] (0.400,0) -- (0.400,0.55);
\fill[blue] (0.400,0.55) circle (1.7pt);
\draw[red!50,very thin] (-0.400,0) -- (-0.400,-0.55);
\fill[red] (-0.400,-0.55) +(-1.7pt,-1.7pt) rectangle +(1.7pt,1.7pt);
\draw[red!50,very thin] (0.900,0) -- (0.900,-0.55);
\fill[red] (0.900,-0.55) +(-1.7pt,-1.7pt) rectangle +(1.7pt,1.7pt);
\draw[red!50,very thin] (-1.400,0) -- (-1.400,-0.55);
\fill[red] (-1.400,-0.55) +(-1.7pt,-1.7pt) rectangle +(1.7pt,1.7pt);
\draw[red!50,very thin] (1.900,0) -- (1.900,-0.55);
\fill[red] (1.900,-0.55) +(-1.7pt,-1.7pt) rectangle +(1.7pt,1.7pt);
\draw[red!50,very thin] (-2.400,0) -- (-2.400,-0.55);
\fill[red] (-2.400,-0.55) +(-1.7pt,-1.7pt) rectangle +(1.7pt,1.7pt);
\draw[red!50,very thin] (2.900,0) -- (2.900,-0.55);
\fill[red] (2.900,-0.55) +(-1.7pt,-1.7pt) rectangle +(1.7pt,1.7pt);
\draw[red!50,very thin] (-3.400,0) -- (-3.400,-0.55);
\fill[red] (-3.400,-0.55) +(-1.7pt,-1.7pt) rectangle +(1.7pt,1.7pt);
\draw (3.400,-0.06) -- (3.400,0.06);
\draw (-2.900,-0.06) -- (-2.900,0.06);
\draw (2.400,-0.06) -- (2.400,0.06);
\draw (-1.900,-0.06) -- (-1.900,0.06);
\draw (1.400,-0.06) -- (1.400,0.06);
\draw (-0.900,-0.06) -- (-0.900,0.06);
\draw (0.400,-0.06) -- (0.400,0.06);
\draw (-0.400,-0.06) -- (-0.400,0.06);
\draw (0.900,-0.06) -- (0.900,0.06);
\draw (-1.400,-0.06) -- (-1.400,0.06);
\draw (1.900,-0.06) -- (1.900,0.06);
\draw (-2.400,-0.06) -- (-2.400,0.06);
\draw (2.900,-0.06) -- (2.900,0.06);
\draw (-3.400,-0.06) -- (-3.400,0.06);
\node[blue,anchor=east] at (-3.750,0.55) {even sublattice};
\node[red,anchor=east] at (-3.750,-0.55) {odd sublattice};
\draw[decorate,decoration={brace,amplitude=3pt}] (0.400,0.78) -- (0.900,0.78) node[midway,above=2pt] {$\tfrac12$};
\draw[decorate,decoration={brace,amplitude=3pt}] (0.900,0.78) -- (1.400,0.78) node[midway,above=2pt] {$\tfrac12$};
\draw[decorate,decoration={brace,mirror,amplitude=3pt}] (-0.400,-0.78) -- (0.400,-0.78) node[midway,below=2pt] {$\tfrac{1+2\mu}{2}$};
\node[anchor=west,font=\small] at (-5.600,1.45) {(b) $N=13$, $\mu=0.3$, $b=0$};
\end{tikzpicture}

\vspace{2pt}

\begin{tikzpicture}[x=1.30cm,y=1cm,font=\scriptsize]
\draw[->,thick] (-4.700,0.45) -- (4.850,0.45);
\fill[blue] (3.400,0.45) circle (1.7pt);
\fill[blue] (-3.100,0.45) circle (1.7pt);
\fill[blue] (2.400,0.45) circle (1.7pt);
\fill[blue] (-2.100,0.45) circle (1.7pt);
\fill[blue] (1.400,0.45) circle (1.7pt);
\fill[blue] (-1.100,0.45) circle (1.7pt);
\fill[blue] (0.400,0.45) circle (1.7pt);
\fill[red] (-0.400,0.45) +(-1.7pt,-1.7pt) rectangle +(1.7pt,1.7pt);
\fill[red] (0.700,0.45) +(-1.7pt,-1.7pt) rectangle +(1.7pt,1.7pt);
\fill[red] (-1.400,0.45) +(-1.7pt,-1.7pt) rectangle +(1.7pt,1.7pt);
\fill[red] (1.700,0.45) +(-1.7pt,-1.7pt) rectangle +(1.7pt,1.7pt);
\fill[red] (-2.400,0.45) +(-1.7pt,-1.7pt) rectangle +(1.7pt,1.7pt);
\fill[red] (2.700,0.45) +(-1.7pt,-1.7pt) rectangle +(1.7pt,1.7pt);
\fill[red] (-3.400,0.45) +(-1.7pt,-1.7pt) rectangle +(1.7pt,1.7pt);
\node[anchor=east] at (-4.750,0.45) {$N=13$};
\draw[->,thick] (-4.700,-0.45) -- (4.850,-0.45);
\fill[blue] (4.400,-0.45) circle (1.7pt);
\fill[blue] (-4.100,-0.45) circle (1.7pt);
\fill[blue] (3.400,-0.45) circle (1.7pt);
\fill[blue] (-3.100,-0.45) circle (1.7pt);
\fill[blue] (2.400,-0.45) circle (1.7pt);
\fill[blue] (-2.100,-0.45) circle (1.7pt);
\fill[blue] (1.400,-0.45) circle (1.7pt);
\fill[blue] (-1.100,-0.45) circle (1.7pt);
\fill[blue] (0.400,-0.45) circle (1.7pt);
\fill[red] (-0.400,-0.45) +(-1.7pt,-1.7pt) rectangle +(1.7pt,1.7pt);
\fill[red] (0.700,-0.45) +(-1.7pt,-1.7pt) rectangle +(1.7pt,1.7pt);
\fill[red] (-1.400,-0.45) +(-1.7pt,-1.7pt) rectangle +(1.7pt,1.7pt);
\fill[red] (1.700,-0.45) +(-1.7pt,-1.7pt) rectangle +(1.7pt,1.7pt);
\fill[red] (-2.400,-0.45) +(-1.7pt,-1.7pt) rectangle +(1.7pt,1.7pt);
\fill[red] (2.700,-0.45) +(-1.7pt,-1.7pt) rectangle +(1.7pt,1.7pt);
\fill[red] (-3.400,-0.45) +(-1.7pt,-1.7pt) rectangle +(1.7pt,1.7pt);
\fill[red] (3.700,-0.45) +(-1.7pt,-1.7pt) rectangle +(1.7pt,1.7pt);
\fill[red] (-4.400,-0.45) +(-1.7pt,-1.7pt) rectangle +(1.7pt,1.7pt);
\node[anchor=east] at (-4.750,-0.45) {$N=17$};
\draw[dashed] (0,-0.85) node[below] {$c_0$} -- (0,0.85);
\draw[gray!60,densely dotted] (-3.400,-0.45) -- (-3.400,0.45);
\draw[gray!60,densely dotted] (-3.100,-0.45) -- (-3.100,0.45);
\draw[gray!60,densely dotted] (-2.400,-0.45) -- (-2.400,0.45);
\draw[gray!60,densely dotted] (-2.100,-0.45) -- (-2.100,0.45);
\draw[gray!60,densely dotted] (-1.400,-0.45) -- (-1.400,0.45);
\draw[gray!60,densely dotted] (-1.100,-0.45) -- (-1.100,0.45);
\draw[gray!60,densely dotted] (-0.400,-0.45) -- (-0.400,0.45);
\draw[gray!60,densely dotted] (0.400,-0.45) -- (0.400,0.45);
\draw[gray!60,densely dotted] (0.700,-0.45) -- (0.700,0.45);
\draw[gray!60,densely dotted] (1.400,-0.45) -- (1.400,0.45);
\draw[gray!60,densely dotted] (1.700,-0.45) -- (1.700,0.45);
\draw[gray!60,densely dotted] (2.400,-0.45) -- (2.400,0.45);
\draw[gray!60,densely dotted] (2.700,-0.45) -- (2.700,0.45);
\draw[gray!60,densely dotted] (3.400,-0.45) -- (3.400,0.45);
\node[anchor=west,font=\small] at (-5.900,1.25) {(c) same $\mu$ and $b$, two sizes $N$: the points at a bounded distance from $c_0$ do not move};
\end{tikzpicture}
\caption{The two--sided bi--lattice \eqref{eq:grid} in terms of $q=x-c_0$, for $N=13$ and
$\mu=0.3$; even sublattice $\{x_{2s}\}$: blue circles, odd sublattice $\{x_{2s+1}\}$: red
squares. (a) $b=0.4$: each sublattice has an arm on each side of the centre $c_0$, each arm an
arithmetic progression of unit step; on each side the two arms interlace, the gaps between
consecutive points alternating between $\tfrac{1-b}{2}$ and $\tfrac{1+b}{2}$, in opposite
orders on the two sides, and the two innermost points $\pm\tfrac{1+2\mu}{4}$ leave the gap
$\tfrac{1+2\mu}{2}$ across the centre. (b) $b=0$: equal spacing $\tfrac12$ on each side and a
grid symmetric about $c_0$, the reflection exchanging the two sublattices. (c) The same $\mu$
and $b$ for $N=13$ and $N=17$: the points at a bounded distance from $c_0$ are the same, which
is why the limit $N\to\infty$ can be taken at fixed $q$.}
\label{fig:grid}
\end{figure}

\paragraph{The shifted operator $Q$.}
The two--sidedness of the grid is the structural novelty of the para--Bannai--Ito Hamiltonian:
the para--Krawtchouk and para--Racah grids are bounded below, and the eigenvalues followed to
the limit sit at their lower end, whereas here they sit around the centre $c_0$. We therefore
set
\begin{equation}
\label{eq:Q}
Q=J-c_0 ,
\end{equation}
whose spectrum straddles $0$; the relevant low--lying states of $Q^2$ --- the states of
bounded level --- sit near the \emph{middle} of the spectrum of $J$, not at an edge, and $Q$ is
a square root of the Schr\"odinger operator $H$. (The letter $q$ will always denote the
eigenvalue $x-c_0$ of $Q$; the base $q$ of the $q$--polynomials of the introduction appears
only in the labels ``$q=-1$'' and ``$q$--Askey scheme''.)

\section{The recurrence as a discrete Dirac equation}
\label{sec:eqlimit}

In terms of the orthonormal polynomials $p_n$ of \S\ref{sec:chain}, the shifted operator $Q$
acts as
\begin{equation}
\label{eq:discrete}
\sqrt{U_n}\,p_{n-1}(x)+(B_n-c_0)\,p_n(x)+\sqrt{U_{n+1}}\,p_{n+1}(x)=q\,p_n(x),\qquad q=x-c_0,
\end{equation}
the degree/site index $n$ playing the role of position and $q$ of (signed) energy. Two features of
\eqref{eq:discrete} distinguish it from the para--Krawtchouk and para--Racah cases and dictate the
nature of the limit: the eigenvalues followed to the limit sit at the centre of the band, not
at an edge, and the couplings are dimerized. We first explain what the first feature means on
a model with constant coefficients, then exhibit both on the recurrence coefficients.
Throughout this section $\alpha=\half$, $N=2j+1$ with $j$ even, $a=-(j+1)-2\mu$ and
$\rho=N/2$.

\paragraph{Band edge and band centre: the Dirac point.}
The mechanism is best seen first on a Jacobi matrix with constant entries, all diagonal entries
equal to $B_\ast$ and all couplings equal to $\kappa_\ast>0$. Its recurrence is solved by the
waves $p_n=e^{\pm\ii kn}$ of wave number (quasimomentum) $k\in[0,\pi]$, with energy
\begin{equation}
\label{eq:band}
x(k)=B_\ast+2\kappa_\ast\cos k ,
\end{equation}
so that the spectrum fills the band $[B_\ast-2\kappa_\ast,B_\ast+2\kappa_\ast]$. Two regimes
must be distinguished. Near an \emph{edge} of the band the wave number is close to $0$ or to
$\pi$: the eigenvectors are slowly varying, or slowly varying up to the alternating sign
$(-1)^n$, and since $x(k)$ is stationary at $k=0$ and $k=\pi$ the energy measured from the edge
is \emph{quadratic} in the wave number, $x-x(k_0)\simeq\mp\kappa_\ast\,(k-k_0)^2$. This is the
dispersion of a Schr\"odinger operator, $(k-k_0)^2$ contracting to $-\partial^2$. The
para--Krawtchouk and para--Racah limits are of this kind \cite{PKcont,PRcont}: their grids are
bounded below, the eigenvalue followed to the limit sits at the bottom of a band whose width
grows with $N$, the eigenvectors are staggered by $(-1)^n$, and the limit is a second--order
operator. Near the \emph{centre} $x=B_\ast$ of the band, on the contrary, $k\simeq\pi/2$ and
\begin{equation}
\label{eq:diracdisp}
x-B_\ast\simeq-2\kappa_\ast\,(k-\tfrac\pi2):
\end{equation}
the dispersion crosses the level $x=B_\ast$ with a non--zero slope, \emph{linearly}. In the
language of band theory this is a \emph{Dirac point}: folded onto the two--site cell, i.e.\
onto the halved Brillouin zone $k\in(-\tfrac\pi2,\tfrac\pi2]$, the band becomes the two bands
$B_\ast\pm2\kappa_\ast\cos k$, which touch at $k=\pm\tfrac\pi2$, and the effective operator
for the nearby energies is of \emph{first order} in the derivative, $k-\tfrac\pi2$
contracting to $-\ii\partial$, as in the Dirac equation. The eigenvectors there carry the fast phase
$e^{\pm\ii\pi n/2}$, whose real forms $\cos\frac{\pi n}2=1,0,-1,0,\dots$ and
$\sin\frac{\pi n}2=0,1,0,-1,\dots$ live respectively on the even and on the odd sites: a real
eigenvector with energy near $B_\ast$ is a slowly varying envelope on the even sites and
another on the odd sites, each multiplied by the period--four sign $(-1)^m$, $n=2m$ or $2m+1$.
The two envelopes are coupled by first differences --- a discrete \emph{Dirac} system --- and
an alternation of the couplings between two values, a \emph{dimerization}, or a staggered
diagonal, enters this system as a mass term. This is the mechanism by which the dimerized
tight--binding model of Su, Schrieffer and Heeger \cite{SSH} is described at low energy by a
Dirac equation whose mass is the dimerization \cite{TLM}, and a model with smoothly varying
couplings by an inhomogeneous Dirac equation with a position--dependent mass \cite{PCLMV}.

\paragraph{The continuum variable.}
We measure positions from the middle of the lattice, $y=n-\rho$ for the site $n$ and
$y=n-\tfrac12-\rho$ for the bond $(n-1,n)$, and we let the site $n$ sit at the position
$x\in(-\tfrac\pi2,\tfrac\pi2)$ defined by
\begin{equation}
\label{eq:x}
\sin x=\frac{2n}{N}-1 ,\qquad\text{i.e.}\qquad y=\rho\sin x,
\end{equation}
which maps the two ends of the lattice to the walls $x=\pm\tfrac\pi2$ and its midpoint
$n=\tfrac N2$ --- the centre of the bond $(j,j+1)$ fixed by $\R_c$ --- to $x=0$. From here on
$x$ denotes this continuum coordinate; the spectral variable will only appear through the grid
points $x_s$. The site reversal $\R_c:n\mapsto N-n$ sends $y$ to $-y$ and, by \eqref{eq:x},
$x$ to $-x$. Let
\begin{equation}
\label{eq:Refl}
\R:\ x\mapsto -x,\qquad \R f(x)=f(-x),
\end{equation}
denote the reflection of functions on the interval.

\begin{figure}[ht]
\centering
\begin{tikzpicture}
\begin{axis}[
  width=0.54\textwidth,height=6.0cm,
  xlabel={$x$},ylabel={$\mathcal V(x)$},
  title={\small generalized P\"oschl--Teller potential},
  xmin=-1.5708,xmax=1.5708,ymin=-2,ymax=8,
  legend style={at={(0.5,0.97)},anchor=north,draw=none,font=\footnotesize},
  tick label style={font=\footnotesize},label style={font=\small},
]
  \addplot[blue,thick,samples=200,domain=-1.45:1.45,restrict y to domain=-3:9]
    {0.25*0.8*(0.8-sin(deg(x)))/(cos(deg(x)))^2};
  \addlegendentry{$\mu=0.8$}
  \addplot[red,thick,samples=200,domain=-1.48:1.48,restrict y to domain=-3:9]
    {0.25*1.4*(1.4-sin(deg(x)))/(cos(deg(x)))^2};
  \addlegendentry{$\mu=1.4$}
\end{axis}
\end{tikzpicture}\hfill
\begin{tikzpicture}
\begin{axis}[
  width=0.44\textwidth,height=6.0cm,
  xlabel={bond $n$},ylabel={$\sqrt{U_n}$},
  title={\small the dimerized couplings ($b=0$)},
  xmin=0,xmax=42,ymin=0,ymax=6.5,
  legend style={at={(0.5,0.02)},anchor=south,draw=none,font=\footnotesize},
  tick label style={font=\footnotesize},label style={font=\small},
]
  \addplot[blue,only marks,mark=*,mark size=1.0pt,samples at={1,3,...,41}]
    {0.25*sqrt((22.6)^2-(x-21)^2)};
  \addlegendentry{odd $n$ (intra--cell)}
  \addplot[red,only marks,mark=square*,mark size=1.0pt,samples at={2,4,...,40}]
    {0.5*sqrt((20-x/2+1)*(x/2))};
  \addlegendentry{even $n$ (inter--cell)}
\end{axis}
\end{tikzpicture}
\caption{Left: the limiting generalized P\"oschl--Teller potential
$\mathcal V(x)=\tfrac14\mu(\mu-\sin x)/\cos^2x$ of \eqref{eq:H}, singular at the two ends
$x=\pm\tfrac\pi2$ (the ends of the lattice), where it behaves as $\tfrac14\mu(\mu\mp1)/(\tfrac\pi2\mp x)^2$:
a repulsive wall at $x=-\tfrac\pi2$, and at $x=+\tfrac\pi2$ a singularity that is attractive
for $\mu<1$ (as for the plotted $\mu=0.8$) and repulsive for $\mu>1$. Right: the couplings
$\sqrt{U_n}$ of $J$
at $b=0$ ($N=41$, $\mu=0.8$), exact from \eqref{eq:A:PR}: the two arcs, the intra--cell couplings
(odd $n$, radius $j+1+2\mu$) and the inter--cell ones (even $n$, radius $j+1$), differ at equal
position by $\mu/(2\cos x)$, a dimerization whose gap is the superpotential, while the diagonal
of $Q$ vanishes identically.}
\label{fig:pot}
\end{figure}
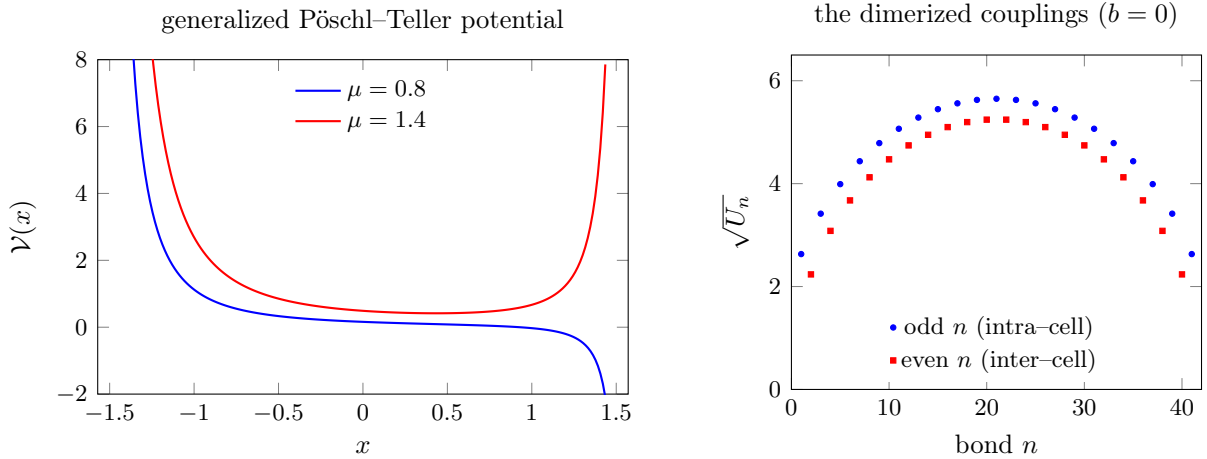

\paragraph{The recurrence coefficients at the persymmetric point.}
At $\alpha=\half$ the two central entries of \eqref{eq:AC} coincide with the generic
formulas, $A_j=\tfrac14(j-j+a)=\tfrac a4$ and $C_{j+1}=-\tfrac14(j+1-j-1-a)=\tfrac a4$, so
that \eqref{eq:AC} holds without exception. Then, for every $b$, the diagonal of $Q$ is
\begin{equation}
\label{eq:A:beta}
\beta_n:=B_n-c_0=\frac{a-j}{4}-A_n-C_n=
\begin{cases}
-\dfrac{nb}{4(n-j-1)}, & n\text{ even},\\[8pt]
\dfrac{(N-n)b}{4(n-j)}, & n\text{ odd},
\end{cases}
\end{equation}
the even case because $\tfrac{a-j}{4}-\tfrac{n-j+a}{4}=-\tfrac n4$ and
$-C_n=\tfrac{n(n-j-1-b)}{4(n-j-1)}=\tfrac n4-\tfrac{nb}{4(n-j-1)}$, the odd case because
$\tfrac{a-j}{4}-C_n=\tfrac{n-2j-1}{4}$ and $A_n=\tfrac{(n-2j-1)(n-j+b)}{4(n-j)}
=\tfrac{n-2j-1}{4}+\tfrac{(n-2j-1)b}{4(n-j)}$. The couplings come in two families, according
to the parity of the bond. We write $u_m=\sqrt{U_{2m+1}}$ for the coupling on the bond
$(2m,2m+1)$, which we call \emph{intra--cell}, and $v_m=\sqrt{U_{2m+2}}$ for the coupling on
the bond $(2m+1,2m+2)$, \emph{inter--cell}, the \emph{cell} $m$ being the pair of sites
$2m,2m+1$. From \eqref{eq:AC},
\begin{align}
u_m^2&=U_{2m+1}=A_{2m}C_{2m+1}=-\tfrac1{16}(2m-j+a)(2m-j-a)
=\tfrac1{16}\big[(j+1+2\mu)^2-(2m-j)^2\big],
\label{eq:A:P}\\
v_m^2&=U_{2m+2}=A_{2m+1}C_{2m+2}
=\frac{(j-m)(m+1)}{4}\cdot\frac{(2m+1-j)^2-b^2}{(2m+1-j)^2},
\label{eq:A:R}
\end{align}
the intra--cell coupling being independent of $b$. In terms of the bond positions
$y=2m+\tfrac12-\rho$ (intra--cell) and $y=2m+\tfrac32-\rho$ (inter--cell), and with
$M_u=j+1+2\mu=|a|$ and $M_v=j+1$, \eqref{eq:A:P}--\eqref{eq:A:R} read
\begin{equation}
\label{eq:A:PR}
u(y)=\tfrac14\sqrt{M_u^2-y^2},\qquad
v(y)=\tfrac14\sqrt{M_v^2-y^2}\;\sqrt{1-b^2/y^2}\,,
\end{equation}
two arcs of the same shape with radii differing by $2\mu$ (Figure~\ref{fig:pot}, right).
Three consequences are immediate. At $b=0$ the diagonal \eqref{eq:A:beta} vanishes
identically: $Q$ is a purely off--diagonal Jacobi matrix, hence \emph{chiral},
$\mathcal C Q\mathcal C=-Q$ with $\mathcal C=\operatorname{diag}((-1)^n)$, which is why its
spectrum is symmetric about $0$ --- the two--sidedness of the grid --- and why the states of
small $|q|$ sit at the centre of the band. The couplings are \emph{dimerized}: the
intra--cell coupling exceeds the inter--cell one by an amount of order one, since by
\eqref{eq:A:PR}, with $y=\rho\sin x$,
\begin{equation}
\label{eq:A:dimer}
u^2-v^2\big|_{b=0}=\tfrac1{16}(M_u^2-M_v^2)=\tfrac14\mu(j+1+\mu),\qquad
u-v=\frac{u^2-v^2}{u+v}=\frac{\mu}{2\cos x}+O(N^{-1}),
\end{equation}
where $u,v=\tfrac14\rho\cos x\,(1+O(N^{-1}))$; in particular, for $x$ away from the walls,
\begin{equation}
\label{eq:Uasym}
\sqrt{U_n}=\frac{j}{4}\cos x\,\big(1+O(N^{-1})\big),
\end{equation}
i.e.\ $\sqrt{U_n}\simeq\tfrac j2\sqrt{(n/N)(1-n/N)}$ in terms of the fraction $n/N$ of the
lattice: the \emph{same} kinetic profile as for the para--Racah Hamiltonian. And for $b\neq0$
the diagonal acquires masses of order one away from the centre: from $n=\rho+y$, $N-n=\rho-y$,
$n-j-1=y-\tfrac12$, $n-j=y+\tfrac12$ and $|y|\gg1$,
\begin{equation}
\label{eq:A:masses}
\beta_{2m}\to\beta_e(x)=-\frac{b(1+\sin x)}{4\sin x},\qquad
\beta_{2m+1}\to\beta_o(x)=\frac{b(1-\sin x)}{4\sin x},
\end{equation}
a common part $-\tfrac b4$ and a staggered part $\mp\tfrac{b}{4\sin x}$, odd in $x$ and
singular at the centre.

\paragraph{$Q$ sits at a Dirac point.}
This is exactly the situation described above. The couplings of $Q$ are of order $N$ in the
bulk \eqref{eq:Uasym}, so that the local band at site $n$, $[-2\sqrt{U_n},2\sqrt{U_n}]$, has a
half--width of order $N$, whereas the eigenvalues $q=x_s-c_0$ that are followed to the limit
are of order one and do not move with $N$ (\S\ref{sec:chain}). They sit at the \emph{centre}
of the band, at the Dirac point, with quasimomentum $\pi/2+O(1/N)$, and their eigenvectors are
not slowly varying but carry the period--four sign $(-1)^{\lfloor n/2\rfloor}$ over slowly
varying envelopes. The dimerization \eqref{eq:A:dimer} will be the superpotential and the
staggered part of the masses \eqref{eq:A:masses} a mass term. Expanding $Q=J-c_0$ about the
Dirac point therefore produces a \emph{first--order} operator, exactly as the two--sidedness
of the grid demands; the second--order Schr\"odinger operator is $H=Q^2$.

\paragraph{An inhomogeneous SSH model.}
At the persymmetric point and $b=0$, then, $Q$ is a chiral Jacobi matrix with dimerized
couplings --- an \emph{inhomogeneous} SSH model in the terminology of \cite{CLMPV}, where such
models are analysed through the \emph{doubling} of orthogonal polynomials \cite{OVdJ1}: a
chiral Jacobi matrix squares to a matrix that is block diagonal on the even and on the odd
sites, each block a Jacobi matrix in its own right, and its polynomials of even degree are
those of the even block in the variable $q^2$ (its polynomials of odd degree, $q$ times those
of the odd block, up to normalization). Here the two blocks are classical. At $b=0$,
\eqref{eq:A:P}--\eqref{eq:A:R} read
$U_{2m+1}=\tfrac14(m+\mu+\tfrac12)(j+\mu+\tfrac12-m)$ and $U_{2m+2}=\tfrac14(m+1)(j-m)$, and
since $Q$ has no diagonal, $Q^2$ has the entries $(Q^2)_{2m,2m}=U_{2m}+U_{2m+1}$,
$(Q^2)_{2m,2m+2}=\sqrt{U_{2m+1}U_{2m+2}}$ on the even sites and
$(Q^2)_{2m+1,2m+1}=U_{2m+1}+U_{2m+2}$, $(Q^2)_{2m+1,2m+3}=\sqrt{U_{2m+2}U_{2m+3}}$ on the odd
ones (with $U_0=U_{N+1}=0$), all entries connecting an even to an odd site vanishing. A
two--line computation,
\begin{align*}
4(Q^2)_{2m,2m}-(\mu+\tfrac12)^2
&=m(j+1-m)+(m+\mu+\tfrac12)(j+\mu+\tfrac12-m)-(\mu+\tfrac12)^2\\
&=(m+\mu+\tfrac12)(j-m)+m(j+\mu+\tfrac32-m),\\
4(Q^2)_{2m+1,2m+1}-(\mu+\tfrac12)^2
&=(m+\mu+\tfrac12)(j+\mu+\tfrac12-m)+(m+1)(j-m)-(\mu+\tfrac12)^2\\
&=(m+\mu+\tfrac32)(j-m)+m(j+\mu+\tfrac12-m),
\end{align*}
with $4(Q^2)_{2m,2m+2}=\sqrt{(m+\mu+\tfrac12)(j-m)(m+1)(j+\mu+\tfrac12-m)}$ and
$4(Q^2)_{2m+1,2m+3}=\sqrt{(m+\mu+\tfrac32)(j-m)(m+1)(j+\mu-\tfrac12-m)}$, then gives
\begin{equation}
\label{eq:dualHahn}
4\,Q^2\big|_{\rm even}=(\mu+\tfrac12)^2\,I+\mathcal J_{\mu-\frac12,\,\mu+\frac12},\qquad
4\,Q^2\big|_{\rm odd}=(\mu+\tfrac12)^2\,I+\mathcal J_{\mu+\frac12,\,\mu-\frac12},
\end{equation}
where $\mathcal J_{\gamma,\delta}$ is the $(j+1)\times(j+1)$ Jacobi matrix of the dual Hahn
polynomials $R_i(\lambda(\ell');\gamma,\delta,j)$ \cite[\S9.6]{KLS}, with diagonal entries
$(i+\gamma+1)(j-i)+i(j+\delta+1-i)$, off--diagonal entries
$\sqrt{(i+\gamma+1)(j-i)(i+1)(j+\delta-i)}$, $i=0,\dots,j$ the row index, and spectrum
$\lambda(\ell')=\ell'(\ell'+\gamma+\delta+1)$, $\ell'=0,\dots,j$ (here $\gamma,\delta$ are the
dual Hahn parameters, used only here); the two blocks are the mirror images of each other, as
the persymmetry of $Q$ requires. At $b=0$ the para--Bannai--Ito
Hamiltonian is thus the double of two dual Hahn families whose parameters differ by units,
$(\gamma,\delta)=(\mu\mp\tfrac12,\mu\pm\tfrac12)$. Its polynomials at $b=0$ are the dual $-1$
Hahn polynomials \cite[\S6.1]{PBI}, which \eqref{eq:dualHahn} exhibits, in the present
normalization, as a double of dual Hahn polynomials; and its spectrum is read off
\eqref{eq:dualHahn}: $q^2=\tfrac14\big[(\mu+\tfrac12)^2+\ell'(\ell'+2\mu+1)\big]
=\tfrac14(\ell'+\mu+\tfrac12)^2$, once per block, i.e.\ $q=\pm\tfrac14(2\ell'+1+2\mu)$,
$\ell'=0,\dots,j$, which is \eqref{eq:towers} at $b=0$, the dual Hahn index $\ell'$ being the
level $\ell$. In this light the continuum limit of
\S\ref{sec:polylimit} is, at $b=0$, the limit of a dual Hahn double taken at fixed $q$; we shall
take it on the para--Bannai--Ito series themselves, for every $b$ --- for $b\neq0$ the diagonal
of $Q$ does not vanish and $Q$ is not a double.

\paragraph{The two--component recast.}
To expose the Dirac structure, split the lattice into even and odd sites. With $N+1=2j+2$ sites,
set, for $m=0,\dots,j$,
\begin{equation}
\label{eq:spinor}
p_{2m}=(-1)^m\varphi_m,\qquad p_{2m+1}=(-1)^m\chi_m,
\end{equation}
which removes the period--four phase; for the states of bounded level, $\varphi$ and $\chi$
will turn out to be slowly varying (this is proved in \S\ref{sec:polylimit}), and we call them
the two \emph{envelopes}. With the $\beta_n$ of \eqref{eq:A:beta}, the single equation
\eqref{eq:discrete} becomes \emph{exactly} the coupled pair
\begin{equation}
\label{eq:dirac-disc}
\begin{aligned}
\beta_{2m}\,\varphi_m+u_m\,\chi_m-v_{m-1}\,\chi_{m-1}&=q\,\varphi_m,\\
u_m\,\varphi_m-v_m\,\varphi_{m+1}+\beta_{2m+1}\,\chi_m&=q\,\chi_m,
\end{aligned}
\end{equation}
(with $v_{-1}=v_j=0$), a discrete \emph{Dirac} system for the two--component field
$(\varphi,\chi)$ --- an exact rewriting of $Qp=qp$. The off--diagonal blocks carry differences of
the couplings and the diagonal blocks carry the masses $\beta_n$, which vanish at $b=0$ (at the
persymmetric point); there is no second difference, hence no Schr\"odinger kinetic term --- $Q$
is first order. This is the discrete Dirac description of inhomogeneous SSH models with smoothly
varying couplings used in \cite{PCLMV}, with $u_m$ and $v_m$ in the roles of the intra-- and
inter--cell hoppings, the cell--wise sign $(-1)^m$ removed in the same way, and the difference
of the two hopping profiles as the mass. There the mass changes sign along the lattice and the
interface carries a Jackiw--Rebbi zero mode \cite{JR} whose extension grows as $\sqrt N$; here,
at $b=0$, the two coupling profiles \eqref{eq:A:PR} satisfy
$u(y)^2-v(y)^2=\tfrac14\mu(j+1+\mu)$ at every bond position $y$ \eqref{eq:A:dimer}, so that the
mass $u-v\to\mu/(2\cos x)$ of the continuum description has, for $\mu>0$, a definite sign; the
spectrum of $Q$ has the gap $\tfrac12(1+2\mu)$ about zero \eqref{eq:towers} and there is no
zero mode: the limit at fixed $q$ resolves instead the bound states of a massive Dirac
operator, which will be the Scarf~I tower.

\paragraph{The mirror on the two--component field.}
At the persymmetric point every eigenvector of $J$ is a mirror eigenvector, $\R_c\,p=\epsilon\,p$
with $\epsilon=\pm1$ (\S\ref{sec:chain}), i.e.\ $p_{N-n}=\epsilon\,p_n$; with $N-2m=2(j-m)+1$
and $(-1)^{j-m}=(-1)^m$ for $j$ even, \eqref{eq:spinor} turns this into the exact relation
$\chi_{j-m}=\epsilon\,\varphi_m$ at finite $N$, and since the sites $2m$ and $2(j-m)+1$ sit at
opposite positions $\pm x$ --- $\sin\xi_{2(j-m)+1}=-\sin\xi_{2m}$ exactly by \eqref{eq:x},
$\xi_n$ denoting the position of the site $n$ --- into
\begin{equation}
\label{eq:channels}
\chi=\epsilon\,\R\varphi
\end{equation}
in the continuum: on the two--component field the site reversal acts as the exchange
\begin{equation}
\label{eq:Mexch}
\mathcal M:\ (\varphi,\chi)\mapsto(\R\chi,\R\varphi),
\end{equation}
and its two eigenspaces, which we call the two \emph{channels}, are the pairs with
$\chi=\pm\R\varphi$ --- on each channel the odd--site
envelope is the reflected even--site envelope. Since $\epsilon=+1$ on the even sublattice and
$-1$ on the odd one \eqref{eq:eps}, the two channels carry the two sublattices of the
spectrum. This is where the reflection of the limiting operator comes from.

\paragraph{The bulk expansion.}
We now expand \eqref{eq:dirac-disc} in the bulk, regarding $\varphi$ and $\chi$ as smooth
functions of $y$, slowly varying on the scale $\rho$, so that each $\partial_y$ lowers the
order by $N$; that the envelopes of the states of bounded level are indeed slowly varying is
the hypothesis of the contraction, established in \S\ref{sec:polylimit}, where the
eigenvectors are shown to converge to the eigenfunctions of the limiting operator. In the
first equation of \eqref{eq:dirac-disc}, written at the site $y_0=2m-\rho$ of $\varphi_m$,
the neighbours are $\chi_m=\chi(y_0+1)$, $\chi_{m-1}=\chi(y_0-1)$, $u_m=u(y_0+\tfrac12)$ and
$v_{m-1}=v(y_0-\tfrac12)$. Expanding,
\begin{equation}
\label{eq:A:expand}
u(y_0+\tfrac12)\chi(y_0+1)-v(y_0-\tfrac12)\chi(y_0-1)
=\big[(u-v)+\tfrac12(u'+v')\big]\chi+(u+v)\,\partial_y\chi+O(N^{-1}),
\end{equation}
all functions being evaluated at $y_0$: the terms $\tfrac12(u'-v')\partial_y\chi$ and
$\tfrac12(u-v)\partial_y^2\chi$ are of order $N^{-1}$ and $N^{-2}$. In the second equation,
written at the site $y_1=2m+1-\rho$ of $\chi_m$, one has $\varphi_m=\varphi(y_1-1)$,
$\varphi_{m+1}=\varphi(y_1+1)$, $u_m=u(y_1-\tfrac12)$, $v_m=v(y_1+\tfrac12)$, and the same
expansion gives $\big[(u-v)-\tfrac12(u'+v')\big]\varphi-(u+v)\,\partial_y\varphi$. Now
$u'(y)=-y/(16u)=-\tfrac14\tan x+O(N^{-1})$ and the same for $v'$, while
$(u+v)\,\partial_y=\tfrac12\rho\cos x\cdot(\rho\cos x)^{-1}\partial_x=\tfrac12\partial_x$
up to $O(N^{-1})$: the variable $x$ is the one in which the velocity is constant, a step of
one site being a step $2/(N\cos x)$ in $x$, so that the sum $u+v\simeq\tfrac j2\cos x$ of two
consecutive couplings, which is what multiplies the first difference of an envelope in
\eqref{eq:dirac-disc}, becomes the constant coefficient $\tfrac12$ of $\partial_x$. With
\eqref{eq:A:dimer}--\eqref{eq:A:masses}, the system \eqref{eq:dirac-disc} contracts to
\begin{equation}
\label{eq:A:dirac0}
q\,\varphi=\beta_e\varphi+\tfrac12\partial_x\chi+\Big(\frac{\mu}{2\cos x}-\frac{\tan x}{4}\Big)\chi,\qquad
q\,\chi=\beta_o\chi-\tfrac12\partial_x\varphi+\Big(\frac{\mu}{2\cos x}+\frac{\tan x}{4}\Big)\varphi :
\end{equation}
the derivative comes from the sum of the two couplings, the superpotential $\mu/\cos x$ from
their difference \eqref{eq:A:dimer}, and the masses from the diagonal.

\paragraph{Normalization.}
The site density is $dn/dx=\rho\cos x$, i.e.\ $dm/dx=\tfrac\rho2\cos x$ for the cells, so that
$\sum_n|p_n|^2=\sum_m(|\varphi_m|^2+|\chi_m|^2)$ becomes
$\tfrac\rho2\int(|\varphi|^2+|\chi|^2)\cos x\,dx$: the amplitudes normalized in $L^2(dx)$ are
\begin{equation}
\label{eq:tilde}
\tilde\varphi=(\cos x)^{1/2}\varphi,\qquad \tilde\chi=(\cos x)^{1/2}\chi ,
\end{equation}
the factor $(\cos x)^{1/2}$ being the square root of the site density. Since
$\tfrac12\partial_x\chi-\tfrac14\tan x\,\chi=(\cos x)^{-1/2}\,\tfrac12\partial_x\big[(\cos x)^{1/2}\chi\big]$,
multiplying \eqref{eq:A:dirac0} by $(\cos x)^{1/2}$ removes the $\tan x$ terms:
\begin{equation}
\label{eq:A:dirac}
q\,\tilde\varphi=\beta_e\,\tilde\varphi+\tfrac12\Big(\partial_x+\frac{\mu}{\cos x}\Big)\tilde\chi,\qquad
q\,\tilde\chi=\beta_o\,\tilde\chi+\tfrac12\Big(-\partial_x+\frac{\mu}{\cos x}\Big)\tilde\varphi .
\end{equation}
This is the limiting form of the discrete Dirac system. The mirror relation \eqref{eq:channels}
reads $\tilde\chi=\epsilon\R\tilde\varphi$ as well, $\cos x$ being even.

\paragraph{The limiting supercharge.}
At $b=0$ the masses vanish and \eqref{eq:A:dirac} is $q\Phi=\mathcal D_\mu\Phi$ with
$\Phi=(\tilde\varphi,\tilde\chi)^{T}$ and
\begin{equation}
\label{eq:A:Dmu}
\mathcal D_\mu=\begin{pmatrix}0&\mathcal A\\ \mathcal A^\dagger&0\end{pmatrix},\qquad
\mathcal A=\tfrac12\Big(\partial_x+\frac{\mu}{\cos x}\Big),\quad
\mathcal A^\dagger=\tfrac12\Big(-\partial_x+\frac{\mu}{\cos x}\Big),
\end{equation}
the supersymmetric pair with superpotential $\mu/\cos x$:
$\mathcal A^\dagger\mathcal A=-\tfrac14\partial_x^2+\tfrac14\mu(\mu-\sin x)/\cos^2x$ acts on
$\tilde\chi$ and $\mathcal A\mathcal A^\dagger$ on $\tilde\varphi$. On substituting the mirror
relation $\tilde\varphi=\epsilon\R\tilde\chi$ in the second equation of \eqref{eq:A:dirac}
and using $\R\,\partial_x\R=-\partial_x$ and $\R\,(\cos x)^{-1}\R=(\cos x)^{-1}$, the
two--component system collapses on each channel to a single first--order operator containing
$\R$,
\begin{equation}
\label{eq:A:channel}
q\,\tilde\chi=\epsilon\,\mathcal A^\dagger\R\,\tilde\chi
=-\epsilon\,\tfrac12\Big(\partial_x-\frac{\mu}{\cos x}\Big)\R\,\tilde\chi=-\epsilon\,\Qs\,\tilde\chi ,
\end{equation}
with the \emph{Dunkl supercharge}
\begin{equation}
\label{eq:supercharge}
\Qs=\tfrac12\Big(\partial_x-\frac{\mu}{\cos x}\Big)\R :
\end{equation}
on the channel $\epsilon=-1$, the odd sublattice, the odd--site envelope is an eigenfunction
of $\Qs$ with eigenvalue $q$, and on the channel $\epsilon=+1$, the even sublattice, of
$-\Qs$; the first equation of \eqref{eq:A:dirac} gives the same statement for the even--site
envelope in mirror image, $q\,\tilde\varphi=-\epsilon\,\R\Qs\R\,\tilde\varphi$. The two
channels thus carry opposite signs of one and the same operator, and
\begin{equation}
\label{eq:H}
H=\Qs^2=-\tfrac14\,\partial_x^2+\frac14\,\frac{\mu(\mu-\sin x)}{\cos^2 x}
\end{equation}
is the generalized P\"oschl--Teller (Scarf~I) Hamiltonian \cite{CKS} on the interval
$(-\tfrac\pi2,\tfrac\pi2)$. The operator \eqref{eq:supercharge} is precisely the ``square root of
the Schr\"odinger operator'' of \cite[Eq.~(7.14)]{VZm1}: with the even superpotential
$W(x)=-\mu/\cos x$, $\Qs=\tfrac12(\partial_x+W)\R$ obeys
$\Qs^2=\tfrac14[-\partial_x^2+W^2+W'] =-\tfrac14\partial_x^2+\tfrac14\mu(\mu-\sin x)/\cos^2x$,
because $W^2+W'=\mu^2/\cos^2x-\mu\sin x/\cos^2x$. The reflection sits inside the
supercharge and \emph{not} in \eqref{eq:H}: $H$ is an ordinary (generalized) P\"oschl--Teller
Hamiltonian --- a pure differential operator, though not a reflection--invariant one, its
potential containing the term $-\tfrac14\mu\sin x/\cos^2x$, odd in $x$ --- whose distinguishing
feature is that it admits a first--order Dunkl square root. Both channels carry the same $H$,
with opposite supercharges $\pm\Qs$; the site reversal $\R_c$, which acts on the two--component
field as the exchange $(\varphi,\chi)\mapsto(\R\chi,\R\varphi)$, is on each channel the number
$\epsilon$ that labels it (\S\ref{sec:reflection}). Figure~\ref{fig:pot} shows the potential
of \eqref{eq:H} and the dimerized couplings that produce it. For $b\neq0$ the masses
\eqref{eq:A:masses} are singular at $x=0$, where the bulk expansion, valid for $|y|\gg1$, does
not apply: as at the central point of the two companions, the layer $|y|=O(1)$ decides the
behaviour of the envelopes at the centre, which the formal expansion of this section cannot
determine, and which the limit taken on the polynomials supplies (\S\ref{sec:polylimit}). The
general case $b\neq0$ is taken up in \S\ref{sec:reflection}.

\begin{remark}
In summary, the mechanism is the mirror image of the two companions. There the relevant states
sat at a band edge, the diagonal was smooth, and the contraction of the second difference
produced a \emph{second--order} Schr\"odinger operator. Here the two--sided grid places them at
the band centre, the diagonal of $Q$ vanishes and the couplings are dimerized, and the
contraction produces a \emph{first--order} Dirac/Dunkl operator whose square is the
Schr\"odinger Hamiltonian. A two--sided spectrum is the spectral signature of a square root.
\end{remark}

\section{The continuum limit taken directly on the polynomials}
\label{sec:polylimit}

We now establish the convergence of the eigenvectors analytically, in parallel with the
confluences Hahn~$\to$~Laguerre and Racah~$\to$~Jacobi of the two companions
\cite{PKcont,PRcont}: the amplitude $p_{2m}(x_s)$ is read as a function of the cell index $m$
at a fixed spectral point $x_s$, where it is a terminating hypergeometric series whose length
does not grow with $N$, and the limit is taken term by term. Throughout this section $m$
labels the cell --- the sites $2m$ and $2m+1$ --- and $\ell$ the level of the spectral point
$x_s$ in its tower \eqref{eq:towers}: $\ell=j-s$ on the even sublattice, $\ell=s$ on the odd
one. The result is the following statement, whose ingredients are assembled in this section
and whose hypergeometric identities are proved in Appendix~\ref{app:confluence}.

\begin{proposition}
\label{prop:conv}
Let $\alpha=\half$, $j$ even, $\mu\ge0$ and $|b|<1$, and fix a grid point $x_s$ of level $\ell$
on the even (respectively odd) sublattice --- a point that does not move with $N$, although
its index $s=j-\ell$ (respectively $s=\ell$) does. For $m=0,\dots,j/2$ let $\xi_m^{(N)}$ be
the position of the site $2m$, $\sin\xi_m^{(N)}=4m/N-1$ by \eqref{eq:x}, and let
$\tilde\varphi_m=(\cos\xi_m^{(N)})^{1/2}(-1)^m\,p_{2m}(x_s)$ be the even--site components of
the eigenvector $\ket{s}$, written as in \eqref{eq:spinor}--\eqref{eq:tilde} and up to the
factor $\sqrt{w_s}$ of its components $\langle e_n|s\rangle=\sqrt{w_s}\,p_n(x_s)$
(\S\ref{sec:chain}). Then there is a constant $\gamma_N$, depending on
$N$ and $s$ but not on $m$, such that for every compact $K\subset(-\tfrac\pi2,0)$
\[
\sup_{m:\ \xi_m^{(N)}\in K}\big|\gamma_N\,\tilde\varphi_m-\Psi_{\ell;\,\pm b,\,2\mu}(-\xi_m^{(N)})\big|
\ \longrightarrow\ 0\qquad(N\to\infty),
\]
with the sign $+b$ on the even sublattice and $-b$ on the odd one; here $\Psi_{\ell;A,B}$ are
the little $-1$ Jacobi functions \eqref{eq:m1wave}. The odd--site components follow from the
exact mirror relation \eqref{eq:channels}, $\tilde\chi(x)=\epsilon\,\tilde\varphi(-x)$, and
the right half of the lattice from the same relation.
\end{proposition}

The convergence is thus established away from the walls and from the centre; the behaviour
of the limit at these points, in particular the exponent $\pm\tfrac b2$ of $|\sin x|$ at the
centre and the exponents $\mu$ of $\cos x$ at $x=\tfrac\pi2$ and $\mu+1$ of $\tfrac\pi2+x$
at $x=-\tfrac\pi2$, is that of the functions $\Psi_{\ell;\pm b,2\mu}$, which are the
eigenfunctions of the supercharges with reflections of \cite{PVZ} with the parameters read
off the grid. This behaviour matters, because the differential expressions alone do not fix
the operators: at $x=\tfrac\pi2$ the Frobenius exponents of $H$ are $\mu$ and $1-\mu$, both
square--integrable when $\mu<\tfrac32$, and at $x=-\tfrac\pi2$ they are $\mu+1$ and $-\mu$,
both square--integrable when $\mu<\tfrac12$, so that boundary conditions are needed there,
as at the centre. Throughout, $\Qs$ and $H$ (and their two--parameter versions
$Q_{\mp b,2\mu}$, $H_{\mp b,2\mu}$ of \S\ref{sec:reflection}) denote the self--adjoint
operators on $L^2(-\tfrac\pi2,\tfrac\pi2)$ diagonal in the complete orthogonal system
$\{\Psi_{\ell;\pm b,2\mu}\}$ (complete because the little $-1$ Jacobi polynomials are a
complete orthogonal system for a positive weight on $[-1,1]$): the self--adjoint
realizations that the limit selects, on whose common domain $\Qs^2=H$ \cite{PVZ}. We first
recall these functions.

\paragraph{The little $-1$ Jacobi functions.}
Recall \cite{VZm1,PVZ} that the little $-1$ Jacobi polynomials $P^{(A,B)}_\ell(\zeta)$ are the
orthogonal polynomials on $[-1,1]$ for the weight $|\zeta|^{A}(1-\zeta^2)^{(B-1)/2}(1+\zeta)$,
$A,B>-1$, and that they satisfy the first--order Dunkl--type eigenvalue equation
\begin{equation}
\label{eq:m1eig}
L_0\,P^{(A,B)}_\ell(\zeta)=\omega_\ell P^{(A,B)}_\ell(\zeta),\qquad
L_0:=2(1-\zeta)\partial_\zeta\R+\big(A+B+1-\tfrac{A}{\zeta}\big)(1-\R),
\end{equation}
with $\omega_\ell=-2\ell$ for $\ell$ even and $\omega_\ell=2(\ell+A+B+1)$ for $\ell$ odd,
where $\R$ is the reflection \eqref{eq:Refl}, which acts as $\zeta\mapsto-\zeta$ on functions
of $\zeta=\sin x$; $L_0$ is the Dunkl operator of \cite{VZm1}, written in our sign convention.
Explicitly \cite[Eqs.~(3.13)--(3.14)]{VZm1}, up to normalization,
\begin{equation}
\label{eq:m1explicit}
\begin{aligned}
P^{(A,B)}_{2k}(\zeta)&\propto{}_2F_1\!\Big(\begin{matrix}-k,\,k+\frac{A+B}2+1\\ \frac{A+1}2\end{matrix};\zeta^2\Big)
+\frac{2k\,\zeta}{A+1}\,{}_2F_1\!\Big(\begin{matrix}1-k,\,k+\frac{A+B}2+1\\ \frac{A+3}2\end{matrix};\zeta^2\Big),\\[3pt]
P^{(A,B)}_{2k-1}(\zeta)&\propto{}_2F_1\!\Big(\begin{matrix}1-k,\,k+\frac{A+B}2\\ \frac{A+1}2\end{matrix};\zeta^2\Big)
-\frac{(A+B+2k)\,\zeta}{A+1}\,{}_2F_1\!\Big(\begin{matrix}1-k,\,k+\frac{A+B}2+1\\ \frac{A+3}2\end{matrix};\zeta^2\Big).
\end{aligned}
\end{equation}
The weighted functions
\begin{equation}
\label{eq:m1wave}
\Psi_{\ell;A,B}(x)=|\sin x|^{A/2}(\cos x)^{B/2}(1+\sin x)^{1/2}\,P^{(A,B)}_\ell(\sin x)
\end{equation}
are, on $(-\tfrac\pi2,\tfrac\pi2)$, the eigenfunctions of the supercharge
\begin{equation}
\label{eq:QAB}
Q_{A,B}=\tfrac1{\sqrt2}\Big(\partial_x-\frac{B}{2\cos x}\Big)\R-\frac{A}{2\sqrt2\,\sin x}
\end{equation}
of \cite[\S4]{PVZ}, with eigenvalues $\tfrac1{2\sqrt2}(-1)^{\ell+1}(2\ell+A+B+1)$, and they are
orthonormal in $L^2(dx)$ up to constants, $\Psi_{\ell;A,B}^2\,dx$ being the little $-1$ Jacobi
weight in the variable $\zeta=\sin x$. In the special case $A=0$ \cite[\S7]{VZm1} the
reflection leaves the potential of the associated Schr\"odinger operator and survives only in
its square root:
$\Psi_{\ell;0,B}(x)\propto(1+\sin x)^{1/2}(\cos x)^{B/2}\,{}_2F_1(-\ell,\ell+B+1;\tfrac{B+1}2;\tfrac{1-\sin x}2)$
are the eigenfunctions of $H_1=-\partial_x^2+\tfrac B2(\tfrac B2-\sin x)/\cos^2x$ with
eigenvalues $(\ell+\tfrac{B+1}2)^2$, and $L_1=(\partial_x-\tfrac{B}{2\cos x})\R$ is its square
root, $L_1^2=H_1$ \cite[Eqs.~(7.14)--(7.18)]{VZm1}; comparing with \eqref{eq:H},
\begin{equation}
\label{eq:id}
H=\tfrac14\,H_1\big|_{B=2\mu},\qquad \Qs=\tfrac12\,L_1\big|_{B=2\mu}=\tfrac1{\sqrt2}\,Q_{0,2\mu}.
\end{equation}

\paragraph{The polynomials.}
The para--Bannai--Ito polynomials of the even case $N'=2j$, which we denote $P'_n$, are given
in \cite[\S3.3]{PBI} as Wilson--type ${}_4F_3$ series, and those of our case $N=2j+1$ as their
Geronimus transforms \cite[\S4.3]{PBI},
\begin{equation}
\label{eq:geronimus}
P_n=P'_n-C_n\,P'_{n-1},
\end{equation}
with the $C_n$ of \eqref{eq:AC}. At $\alpha=\half$, with $a=-(j+1)-2\mu$ and the abbreviation
\begin{equation}
\label{eq:mb}
m_b=\frac{2\mu+b}4=\frac{b-j-1-a}4,
\end{equation}
the series of \cite[\S3.3]{PBI} read, at a grid point $x_s$ and on the left half of the
lattice ($2m\le j$ for the even degrees, $2m+1<j$ for the odd ones),
\begin{equation}
\label{eq:F43}
P'_{2m}(x_s)=\varkappa_m\,F_m(x_s),\qquad P'_{2m+1}(x_s)=\varkappa'_m\,(x_s-m_b)\,G_m(x_s),
\end{equation}
\begin{equation}
\label{eq:F43series}
\begin{aligned}
F_m(x_s)&={}_4F_3\!\left(\begin{matrix}-m,\ m-j,\ m_b+x_s,\ m_b-x_s\\
\tfrac{1+b-j}2,\ \mu+\tfrac12,\ -\tfrac j2\end{matrix};1\right),\\[3pt]
G_m(x_s)&={}_4F_3\!\left(\begin{matrix}-m,\ m+1-j,\ m_b+1+x_s,\ m_b+1-x_s\\
\tfrac{3+b-j}2,\ \mu+\tfrac32,\ 1-\tfrac j2\end{matrix};1\right),
\end{aligned}
\end{equation}
with the monic normalizations
$\varkappa_m=(\tfrac{1+b-j}2)_m(\mu+\tfrac12)_m(-\tfrac j2)_m/(m-j)_m$ and
$\varkappa'_m=(\tfrac{3+b-j}2)_m(\mu+\tfrac32)_m(1-\tfrac j2)_m/(m+1-j)_m$ (on the right half the
series of \cite{PBI} acquire a second term, which we shall not need: that half of the lattice
will be reached through the mirror). The sign of $\varkappa_m$ is $(-1)^m$, which is the
staggering \eqref{eq:spinor}. Note that $m_b\pm x_s$ are the arguments in which the grid is
arithmetic: by \eqref{eq:towers}, the even sublattice $\{x_{2s}\}$ consists of the points
\begin{equation}
\label{eq:towerpts}
x_s=m_b+n'\ \ (n'\ge0,\ \ell=2n'),\qquad x_s=-(m_b+n')\ \ (n'\ge1,\ \ell=2n'-1),
\end{equation}
the two signs being its two arms (\S\ref{sec:chain}), and the odd sublattice $\{x_{2s+1}\}$
consists of the points $x_s=m'_b+n'$ ($n'\ge0$, $\ell=2n'+1$) and $x_s=-(m'_b+n')$ ($n'\ge0$,
$\ell=2n'$), with
\begin{equation}
\label{eq:mbprime}
m'_b=\frac{2+2\mu-b}4=\mu+\tfrac12-m_b .
\end{equation}

\paragraph{Duality.}
At a point \eqref{eq:towerpts} of the even sublattice the factor $(m_b\mp x_s)_k=(-n')_k$
terminates $F_m$ at $k=n'$, and $(m_b+1\mp x_s)_k=(1-n')_k$ terminates $G_m$ at $k=n'-1$ (for
$n'=0$ the prefactor $x_s-m_b$ of $P'_{2m+1}$ vanishes instead), \emph{whatever $m$}: as
functions of the cell index the amplitudes are terminating series of fixed length, i.e.\
polynomials in $m(m-j)$, exactly as the Hahn and Racah amplitudes of the companions are
polynomials in $n(n-N)$. This is the duality that makes the limit computable, and the odd
sublattice has it too: the series \eqref{eq:F43series} are balanced, and Whipple's
transformation of a terminating balanced ${}_4F_3$ \cite{Bailey} rewrites them, up to a factor
$\Pi_m$ independent of $x_s$, with the pair $(m'_b\mp x_s)$ in place of $(m_b\mp x_s)$
(Appendix~\ref{app:C1}), which terminates at the points $\pm(m'_b+n')$.

\paragraph{The confluence.}
Put the site $2m$ at the position $x$, i.e.\ $2m=\rho(1+\sin x)$ with $\rho=N/2$, as in
\eqref{eq:x}. Then
\begin{equation}
\label{eq:termlimit}
\frac{(-m)_k\,(m-j)_k}{(\tfrac{1+b-j}2)_k\,(-\tfrac j2)_k}
=\prod_{i=0}^{k-1}\frac{(m-i)(j-m-i)}{(\tfrac{j-1-b}2-i)(\tfrac j2-i)}
=\Big(\frac{4m(j-m)}{j^2}\Big)^{k}\big(1+O(N^{-1})\big)\ \longrightarrow\ (\cos^2x)^k,
\end{equation}
so that the terminating series tend, term by term, to Gauss series in $\cos^2x$: at
$x_s=\pm(m_b+n')$,
\begin{equation}
\label{eq:FGlimit}
\begin{aligned}
F_m(x_s)&\to F(x)={}_2F_1\!\Big(\begin{matrix}-n',\ n'+2m_b\\ \mu+\tfrac12\end{matrix};\cos^2x\Big),\\
G_{m}(x_s)&\to G(x)={}_2F_1\!\Big(\begin{matrix}1-n',\ n'+1+2m_b\\ \mu+\tfrac32\end{matrix};\cos^2x\Big),
\end{aligned}
\end{equation}
and likewise on the odd sublattice with $m_b\to m'_b$ and the parameters of the transformed
series. These are Jacobi polynomials in $\cos2x$; the sign of $x_s$ has disappeared from
them, and will reappear through \eqref{eq:geronimus}.

\paragraph{The Geronimus combination and the profile.}
On the even sites, \eqref{eq:geronimus} gives
$P_{2m}(x_s)=\varkappa_m\big[F_m(x_s)-C_{2m}\tfrac{\varkappa'_{m-1}}{\varkappa_m}(x_s-m_b)\,G_{m-1}(x_s)\big]$,
and the coefficient has a finite limit (Appendix~\ref{app:C2}),
\begin{equation}
\label{eq:coeflimit}
-\,C_{2m}\,\frac{\varkappa'_{m-1}}{\varkappa_m}\ \longrightarrow\ c(x)=\frac{2(1+\sin x)\sin x}{2\mu+1},
\end{equation}
the two terms of the Geronimus combination being of the same order --- the odd--degree
polynomial, smaller by a factor of order $N$, is compensated by $C_{2m}$, of order $N$. The
component of the normalized eigenvector is $\sqrt{w_s}\,P_{2m}(x_s)/h_{2m}$, with
$h_{2m}=\sqrt{U_1\cdots U_{2m}}$, and the factor $\varkappa_m/h_{2m}$, common to all $s$, is the
\emph{profile} of the even sites. From the closed form of $h_{2m}$ \cite[\S4.4]{PBI},
\begin{equation}
\label{eq:envelopeexact}
\Big(\frac{\varkappa_m}{h_{2m}}\Big)^2=\frac{(-j)_m\,(\tfrac{1+b-j}2)_m\,(\mu+\tfrac12)_m}
{m!\,(\tfrac{1-b-j}2)_m\,(-j-\tfrac12-\mu)_m},
\end{equation}
which Stirling's formula evaluates (Appendix~\ref{app:C3}) as
$m^{\mu-\frac12}(j-m)^{\mu+\frac12}\,|\tfrac{j+1}2-m|^{b}$ up to a constant depending on $N$,
whence, with $2m=\rho(1+\sin x)$,
\begin{equation}
\label{eq:envelope}
(-1)^m\,\frac{\varkappa_m}{h_{2m}}\ \simeq\ c_N\,(\cos x)^{\mu-\frac12}\,|\sin x|^{b/2}\,(1-\sin x)^{1/2},
\end{equation}
where $\simeq$ denotes equality up to a factor $1+O(N^{-1})$, uniformly on compact subsets of
$(-\tfrac\pi2,0)$, and $c_N>0$ depends on $N$ only (the sign $(-1)^m$ of $\varkappa_m$ has been
taken out). (This profile is $\Psi_{0;b,2\mu}(-x)$, up to
$(\cos x)^{1/2}$, and for good reason: at $x_s=m_b$, the level $\ell=0$ of the even sublattice,
every $k\ge1$ term of $F_m$ vanishes and $G_{m-1}$ is killed by $x_s-m_b=0$, so that
$P_{2m}(m_b)=\varkappa_m$ exactly and the profile \emph{is} the even--site part of that
eigenvector.)

\paragraph{Result.}
Collecting \eqref{eq:FGlimit}--\eqref{eq:envelope}, the even--site components of the
eigenvector of $x_s=\pm(m_b+n')$ satisfy, on the left half,
\begin{equation}
\label{eq:limitform}
c_N^{-1}\,(-1)^m p_{2m}(x_s)\ \longrightarrow\
(\cos x)^{\mu-\frac12}\,|\sin x|^{b/2}\,(1-\sin x)^{1/2}\,\Big[F(x)+c(x)\,(x_s-m_b)\,G(x)\Big],
\end{equation}
with $x_s-m_b=n'$ or $-(2m_b+n')$ according to the sign. The bracket is a little $-1$ Jacobi
polynomial in disguise: the connection formula $z\to1-z$ of a terminating Gauss series turns
$F$ and $G$ into series in $\sin^2x$ with the lower parameters $\tfrac{b+1}2$ and
$\tfrac{b+3}2$, and two contiguous relations of the ${}_2F_1$
(Appendix~\ref{app:C4}) reassemble the bracket into \eqref{eq:m1explicit}:
\begin{equation}
\label{eq:bracket}
F(x)+c(x)\,(x_s-m_b)\,G(x)\propto P^{(b,\,2\mu)}_{\ell}(-\sin x),\qquad
\ell=\begin{cases}2n',&x_s=m_b+n',\\ 2n'-1,&x_s=-(m_b+n'),\end{cases}
\end{equation}
which is the level $\ell=j-s$ of \eqref{eq:towers}. Hence, restoring the $L^2(dx)$
normalization $(\cos x)^{1/2}$ of \eqref{eq:tilde}, the even--site envelope of the eigenvector
belonging to $x_{2s}$ tends to
\begin{equation}
\label{eq:evenlimit}
\tilde\varphi(x)\ \propto\ \Psi_{j-s;\,b,\,2\mu}(-x)
=|\sin x|^{b/2}(\cos x)^{\mu}(1-\sin x)^{1/2}\,P^{(b,2\mu)}_{j-s}(-\sin x):
\end{equation}
the reflected little $-1$ Jacobi function with parameters $(b,2\mu)$. On the odd sublattice,
$x_s=\pm(m'_b+n')$, the same steps are taken on the Whipple forms $F_m=\Pi_mF^W_m$,
$G_m=\Pi'_mG^W_m$ of Appendix~\ref{app:C1} (valid for $2m<j$, i.e.\ away from the central
cell): factoring $\Pi_m$ out of the Geronimus combination, which involves $G_{m-1}$, its
coefficient acquires the ratio $\Pi'_{m-1}/\Pi_m=(\Pi'_{m-1}/\Pi_{m-1})(\Pi_{m-1}/\Pi_m)$,
whose two factors tend to $1/\sin^2x$ and to $1$ by \eqref{eq:C:Pi}, so that the coefficient
tends to $c(x)/\sin^2x$; the series tend to $F^W={}_2F_1(-n',n'+2m'_b;\mu+\tfrac12;\cos^2x)$ and
$G^W={}_2F_1(-n',n'+2m'_b;\mu+\tfrac32;\cos^2x)$, and the profile acquires the factor
$\Pi_m\propto|\sin x|^{1-b}$, which turns its exponent $\tfrac b2$ at the centre into
$1-\tfrac b2$. The identity
\begin{equation}
\label{eq:bracketW}
\sin x\,F^W(x)+\frac{2(1+\sin x)}{2\mu+1}\,(x_s-m_b)\,G^W(x)\propto P^{(-b,\,2\mu)}_{\ell}(-\sin x),
\end{equation}
with $\ell=2n'+1$ for $x_s=m'_b+n'$ and $\ell=2n'$ for $x_s=-(m'_b+n')$
(Appendix~\ref{app:C4}), then gives $\tilde\varphi\propto\Psi_{\ell;-b,2\mu}(-x)$ with
$\ell=s$, the level of \eqref{eq:towers}: the factor $1/\sin x$ that \eqref{eq:bracketW}
leaves in front of the little $-1$ Jacobi polynomial brings the exponent at the centre from
$1-\tfrac b2$ back to $-\tfrac b2$, that of $\Psi_{\ell;-b,2\mu}$, and the parameters are
$(-b,2\mu)$. The odd sites are obtained in the same way from
$P_{2m+1}=P'_{2m+1}-C_{2m+1}P'_{2m}$: the coefficient $C_{2m+1}\varkappa_m/\varkappa'_m$ tends to
$-1/c(-x)$, the bracket becomes its value at $-x$, and the profile becomes
$(\cos x)^{\mu-\frac12}|\sin x|^{b/2}(1+\sin x)^{1/2}$ with the same constant
(Appendix~\ref{app:C3}), so that $\tilde\chi\propto\Psi_{\ell;\pm b,2\mu}(x)$ --- the mirror
image, as the exact relation \eqref{eq:channels} demands --- and the same relation carries
both envelopes to the right half of the lattice. This proves Proposition~\ref{prop:conv}.

For each fixed level, the eigenvector of $J$ thus converges, with the parameters $(\pm b,2\mu)$
on the two sublattices, to the eigenfunctions of the supercharges $Q_{\pm b,2\mu}$ of
\eqref{eq:QAB}; the corresponding statement for the operators is \eqref{eq:limitb} of the
next section, and the spectrum \eqref{eq:towers} is the spectrum of those supercharges. At
$b=0$ the two sublattices carry the same functions and, by \eqref{eq:id}, the eigenfunctions
of $H=\Qs^2$ are $\Psi_{\ell;0,2\mu}$, with
\begin{equation}
\label{eq:spec}
H\,\Psi_\ell=\tfrac14\big(\ell+\mu+\tfrac12\big)^2\Psi_\ell,\qquad
\Qs\,\Psi_\ell=(-1)^{\ell+1}\tfrac12\big(\ell+\mu+\tfrac12\big)\Psi_\ell,\qquad \ell=0,1,2,\dots,
\end{equation}
the supercharge eigenvalues alternating in sign with $\ell$ --- the two--sided spectrum of
\S\ref{sec:chain}. Equation \eqref{eq:spec} agrees with the exact grid \eqref{eq:towers}: at
$b=0$ the values of $|Q|$ on the grid are $\tfrac14(2\ell+1+2\mu)$, $\ell=0,1,\dots$, each
appearing twice (once per sublattice), and their squares are $\tfrac14(\ell+\mu+\tfrac12)^2$.
The ground state $\Psi_0=(1+\sin x)^{1/2}(\cos x)^{\mu}$ is the square root of the
orthogonality measure, $\Psi_0^2\,dx$ being the weight of the $P^{(0,2\mu)}_\ell$ in the
variable $x$. The para--Bannai--Ito Hamiltonian, shifted to $Q=J-c_0$, thus contracts to the
Dunkl supercharge \eqref{eq:supercharge}, and $H=Q^2$ is the generalized P\"oschl--Teller
Hamiltonian \eqref{eq:H}, whose eigenfunctions are the weighted little $-1$ Jacobi functions
and whose spectrum \eqref{eq:spec} is quadratic in $\ell$ yet borne by a first--order square
root. For $b\neq0$ the two sublattices carry distinct parameters and the limit is the
two--parameter extended Scarf~I Hamiltonian with reflection of \cite{PVZ}; we turn to this next.

\section{Supersymmetry with reflections and the two channels}
\label{sec:reflection}

Equations \eqref{eq:supercharge}--\eqref{eq:H} are an instance of the supersymmetric quantum
mechanics \emph{with reflections} of \cite{PVZ}, of the chiral type sometimes written
$\mathcal N=\tfrac12$: a single Hermitian supercharge whose square is the Hamiltonian. In that
construction a supercharge
\begin{equation}
\label{eq:postQ}
Q[\mathsf U,\mathsf V]=\tfrac1{\sqrt2}\big(\partial_x+\mathsf U(x)\big)\R+\tfrac1{\sqrt2}\mathsf V(x),\qquad \mathsf U\text{ even},\ \mathsf V\text{ odd},
\end{equation}
is Hermitian and yields
$Q[\mathsf U,\mathsf V]^2=-\tfrac12\partial_x^2+\tfrac12(\mathsf U^2+\mathsf V^2)+\tfrac12\mathsf U'-\tfrac12\mathsf V'\R$,
so that the reflection appears in the Hamiltonian unless $\mathsf V$ is constant (the
sans--serif $\mathsf U$, $\mathsf V$ are functions, not to be confused with the recurrence
coefficients $U_n$). The choice $\mathsf U=-B/(2\cos x)$,
$\mathsf V=-A/(2\sin x)$ is the supercharge $Q_{A,B}$ of \eqref{eq:QAB}, and $H_{A,B}=Q_{A,B}^2$ is the
\emph{extended Scarf~I} Hamiltonian, exactly solvable with the little $-1$ Jacobi
eigenfunctions $\Psi_{\ell;A,B}$ and spectrum $(2\ell+A+B+1)^2/8$ \cite[\S4]{PVZ}. Our $b=0$
limit is the $A=0$, $B=2\mu$ member of this family, \eqref{eq:id}: the reflection is carried
by the supercharge and is absent from the Hamiltonian as an operator (\S\ref{sec:eqlimit}).

\paragraph{The second parameter and the two channels.}
At $b=0$ the two channels of \S\ref{sec:eqlimit} --- the two mirror eigenspaces
\eqref{eq:channels} --- carry the opposite supercharges $\pm\Qs$ and the \emph{same}
Hamiltonian $H=\Qs^2$, with the same eigenfunctions: the two sublattices of
\S\ref{sec:polylimit} carry the same functions $\Psi_{\ell;0,2\mu}$, so that the even--site
envelopes of the eigenvectors with eigenvalues $\pm q$ coincide. On a single channel the spectrum of $H$
is simple, the tower \eqref{eq:spec}; the doubling seen on the grid, each $|Q|$ appearing twice
at $b=0$, is the pair of channels, i.e.\ the pair of eigenstates of opposite supercharge
eigenvalue $\pm\tfrac12(\ell+\mu+\tfrac12)$, which the mirror $\R_c$ tells apart by their sign
$\epsilon$.

The remaining grid parameter $b$ enters the limit as the odd superpotential $\mathsf V$ of
\eqref{eq:postQ}, hence as a reflection term in the Hamiltonian. For $b\neq0$ the diagonal of
$Q$ acquires the masses \eqref{eq:A:masses}, a common shift $-b/4$ and a staggered part
$\mp b/(4\sin x)$. With these masses, the substitution $\tilde\varphi=\epsilon\R\tilde\chi$ in the second equation of
\eqref{eq:A:dirac} gives, on the odd--site envelope of the channel $\epsilon$,
\begin{equation}
\label{eq:A:channelb}
\Big(q+\frac b4\Big)\tilde\chi=\Big(-\epsilon\,\Qs+\frac{b}{4\sin x}\Big)\tilde\chi :
\end{equation}
the shift means that the operator with a pure Dunkl limit is $J+\tfrac14=Q+\tfrac b4$ rather
than $Q$, and the staggered part is exactly an odd superpotential $\mathsf V$. In terms of the
supercharge $Q_{A,B}$ of \eqref{eq:QAB}, by \eqref{eq:id}, \eqref{eq:A:channelb} reads
\begin{equation}
\label{eq:limitb}
J+\tfrac14\ \longrightarrow\
\begin{cases}
\phantom{-}\tfrac1{\sqrt2}\,Q_{-b,\,2\mu}, & \text{odd sublattice }(\epsilon=-1),\\[3pt]
-\tfrac1{\sqrt2}\,Q_{\,b,\,2\mu}, & \text{even sublattice }(\epsilon=+1),
\end{cases}
\end{equation}
acting on the odd--site envelope. The eigenvalues of $Q_{A,B}$ are
$\tfrac1{2\sqrt2}(-1)^{\ell+1}(2\ell+A+B+1)$, $\ell=0,1,\dots$ \cite[Eq.~(4.21)]{PVZ}, so
\eqref{eq:limitb} predicts the spectrum $(-1)^{\ell+1}\tfrac14(2\ell+1+2\mu-b)$ on the odd
channel and $(-1)^{\ell}\tfrac14(2\ell+1+2\mu+b)$ on the even one: these are precisely the
grid values $x_{2s+1}+\tfrac14$ (with $\ell=s$) and $x_{2s}+\tfrac14$ (with $\ell=j-s$) of
\eqref{eq:towers}. Hence $(J+\tfrac14)^2\to\tfrac12H_{A,2\mu}$ with $A=-b$ on
the odd and $A=b$ on the even sublattice, where
\begin{equation}
\label{eq:Hb}
\tfrac12H_{A,2\mu}=-\tfrac14\partial_x^2+\frac{\mu(\mu-\sin x)}{4\cos^2x}
+\frac{A^2}{16\sin^2x}-\frac{A\cos x}{8\sin^2x}\,\R
\end{equation}
is the two--parameter extended Scarf~I Hamiltonian with reflection of \cite[Eq.~(4.4)]{PVZ}:
the second parameter adds to the Hamiltonian a term proportional to $\R$, with opposite signs
on the two channels, and a centrifugal term at the centre (the little $-1$ Jacobi weight
requires $A>-1$, i.e.\ $|b|<1$, which is the range \eqref{eq:pos}). Its eigenfunctions are
the two--parameter little $-1$ Jacobi functions $\Psi_{\ell;A,2\mu}$ of \eqref{eq:m1wave}
\cite{PVZ} --- precisely the limits of the odd--site envelopes of the eigenvectors found in
\S\ref{sec:polylimit}, $\tilde\chi\propto\Psi_{\ell;\pm b,2\mu}(x)$, sublattice by sublattice
(the even--site envelopes, their mirror images, are the eigenfunctions of $\R H_{A,2\mu}\R$)
--- and its spectrum is the grid itself, as just noted. At $b=0$ the two towers coincide and
every level of $H$ is doubly degenerate, the pair of channels; for $b\neq0$ they are displaced
by $\mp b/4$, the lifting of the degeneracy by the second parameter. The operators
$Q_{\mp b,2\mu}$ have at $x=0$ a regular singular point, where the masses are singular and
the bulk expansion of \S\ref{sec:eqlimit} does not apply; the little $-1$ Jacobi
eigenfunctions $\Psi_{\ell;A,B}$ select one of its self--adjoint extensions \cite{PVZ}, and
Proposition~\ref{prop:conv} shows that the eigenvectors converge to these functions,
sublattice by sublattice, uniformly on compact subsets away from the centre, with the
exponents $|\sin x|^{\pm b/2}$ of these functions at the centre, which identifies the extension
selected by the discrete model with that of \cite{PVZ}. In the language of the
two companions: the para--Bannai--Ito Hamiltonian consists of two little $-1$ Jacobi channels
--- the two grid sublattices --- which the mirror $\R_c$ combines into a single supersymmetric
system whose supercharge is a Dunkl operator.

\section{The other parity classes: even \texorpdfstring{$N$}{N} and matrix supersymmetry}
\label{sec:even}

The para--Bannai--Ito polynomials come in four classes, according to the residue of $N$
modulo $4$ \cite{PBI}: $N=2j+1$ with $j$ even, the class treated so far, $N=2j+1$ with $j$
odd \cite[App.~B]{PBI}, and $N=2j$ with $j$ even \cite[\S3]{PBI} or odd \cite[App.~A]{PBI}.
All four are persymmetric at $\alpha=\half$, and with the uniform parametrization
$a=-(j+1)-2\mu$ all four have grids given by the formulas \eqref{eq:towers}, with the sign of
every point reversed when $j$ is odd, and with the odd sublattice stopping at the level
$\ell=j-1$ when $N$ is even. The parity of $j$ turns out to be immaterial: for odd $N$ it
reverses the sign of the limiting operator, for even $N$ the sign of $b$, and in both cases
it reverses the sign with which the site reversal acts on the two--component field. The
parity of $N$ is not: it changes the limiting operator. This section establishes both
statements; the computations are those of \S\ref{sec:eqlimit} with different lattice data.

\paragraph{Odd $N$, odd $j$.}
For $N=2j+1$ with $j$ odd the coefficients of \cite[App.~B]{PBI} are those of \eqref{eq:AC}
with the $b$--dependent factors moved from one parity of $n$ to the other,
\begin{equation}
\label{eq:ACodd}
\begin{aligned}
A_n&=\begin{cases}
\tfrac14\dfrac{(n-j+a)(n-j+b)}{n-j}, & n\text{ even},\\[6pt]
\tfrac14(n-2j-1), & n\text{ odd},\ n\neq j,\\[2pt]
-\tfrac12\,\alpha\,(j+1), & n=j,
\end{cases}
\\[4pt]
C_n&=\begin{cases}
-\tfrac14\,n, & n\text{ even},\ n\neq j+1,\\[2pt]
-\tfrac14\dfrac{(n-j-1-a)(n-j-1-b)}{n-j-1}, & n\text{ odd},\\[6pt]
-\tfrac12(1-\alpha)(j+1), & n=j+1,
\end{cases}
\end{aligned}
\end{equation}
and our parametrization $a=-(j+1)-2\mu$, $\mu\ge0$, again gives $U_n>0$ for $n=1,\dots,N$
when $|b|<1$ and $0<\alpha<1$ (from \eqref{eq:ACodd}, $U_n=\tfrac1{16}n(N+1-n)$ for even
$n\neq j+1$, $U_{j+1}=\tfrac14\alpha(1-\alpha)(j+1)^2$, and
$U_n=\tfrac1{16}\,[a^2-(n-j-1)^2]\,[(n-j-1)^2-b^2]/(n-j-1)^2$ for odd $n$, where $n-j-1$
is odd). At $\alpha=\half$ the two central entries again agree with
the generic formulas, and the resulting $B_n$ and $U_n$ are, at $b=0$, given by the same
expressions as for $j$ even, so that nothing changes at all. For
$b\neq0$ two things move: the masses \eqref{eq:A:masses} are exchanged between the even and
the odd sites, up to $O(N^{-1})$, and the factor $\sqrt{1-b^2/y^2}$ of \eqref{eq:A:PR} passes
from the inter--cell to the intra--cell coupling, which does not affect the bulk expansion of
\S\ref{sec:eqlimit}. The grid is \eqref{eq:towers} with the sign of every point reversed,
\begin{equation}
\label{eq:towersodd}
x_{2s}+\tfrac14=(-1)^{\ell+1}\,\frac{2\ell+1+2\mu+b}{4}\ \ (\ell=j-s),\qquad
x_{2s+1}+\tfrac14=(-1)^{\ell}\,\frac{2\ell+1+2\mu-b}{4}\ \ (\ell=s),
\end{equation}
still centred at $c_0$ (as a set of points, \eqref{eq:towersodd} is \eqref{eq:towers} with
$b\to-b$ and the two sublattices exchanged). The largest eigenvalue is still $x_0$, on the even
sublattice, so that $\epsilon=+1$ on $\{x_{2s}\}$ and $-1$ on $\{x_{2s+1}\}$ as before. What
changes is the pairing between the channel sign and the two envelopes: the staggering
\eqref{eq:spinor} acquires the sign $(-1)^j$ under the mirror, so that the exact relation
$\chi_{j-m}=\epsilon\varphi_m$ of \S\ref{sec:eqlimit} becomes $\chi_{j-m}=-\epsilon\varphi_m$. In
the contraction of \S\ref{sec:reflection} the two changes --- the sign of $\epsilon$ in
\eqref{eq:A:channelb} and the sign of the mass on the odd sites --- combine into an overall
sign: on the odd--site envelope of each channel,
\begin{equation}
\label{eq:limitbodd}
J+\tfrac14\ \longrightarrow\
\begin{cases}
\phantom{-}\tfrac1{\sqrt2}\,Q_{\,b,\,2\mu}, & \text{even sublattice }(\epsilon=+1),\\[3pt]
-\tfrac1{\sqrt2}\,Q_{-b,\,2\mu}, & \text{odd sublattice }(\epsilon=-1),
\end{cases}
\end{equation}
the negative of \eqref{eq:limitb} channel by channel --- the same parameters $A=\pm b$ on the
same sublattices, the eigenvalues reversed, in agreement with \eqref{eq:towersodd}. The
convergence of the eigenvectors, Proposition~\ref{prop:conv}, holds as well: the polynomials
of \cite[App.~A--B]{PBI} have the form \eqref{eq:geronimus}--\eqref{eq:F43series} with $m_b$
replaced by $m_{-b}$ and the lower parameters $\tfrac{1+b-j}2$, $-\tfrac j2$ replaced by
$-\tfrac{b+j}2$, $\tfrac{1-j}2$, so that the untransformed series now terminate on the odd
sublattice and their Whipple transforms on the even one; the norm of \cite[App.~B]{PBI} gives
for $(\varkappa_m/h_{2m})^2$ the expression \eqref{eq:envelopeexact} with $\tfrac{1+b-j}2$,
$\tfrac{1-b-j}2$ replaced by $-\tfrac{b+j}2$, $\tfrac{b-j}2$, whose exponent at the centre is
$-\tfrac b2$ instead of $\tfrac b2$; and the
steps of \S\ref{sec:polylimit}, whose limits are insensitive to the half--unit shifts of the
parameters, apply with $b\to-b$ and the two sublattices exchanged --- which is
\eqref{eq:limitbodd}. Everything in \S\S\ref{sec:eqlimit}--\ref{sec:reflection} therefore holds for
$j$ odd with $J+\tfrac14$ replaced by $-(J+\tfrac14)$, i.e.\ with the signs of the two
supercharges reversed; what this does to state transfer is recorded at the end of
\S\ref{sec:FR}.

\begin{figure}[ht]
\centering
\begin{tikzpicture}
\begin{axis}[
  width=0.5\textwidth,height=6.0cm,
  xlabel={bond $n$},ylabel={$\sqrt{U_n}$},
  title={\small the couplings for even $N$ ($b=0$)},
  xmin=0,xmax=41,ymin=0,ymax=6.5,
  legend style={at={(0.5,0.02)},anchor=south,draw=none,font=\footnotesize},
  tick label style={font=\footnotesize},label style={font=\small},
]
  \addplot[blue,only marks,mark=*,mark size=1.0pt,samples at={1,3,...,39}]
    {0.25*sqrt((21.3)^2-(x-20.5+0.8)^2)};
  \addlegendentry{odd $n$}
  \addplot[red,only marks,mark=square*,mark size=1.0pt,samples at={2,4,...,40}]
    {0.25*sqrt((21.3)^2-(x-20.5-0.8)^2)};
  \addlegendentry{even $n$}
\end{axis}
\end{tikzpicture}
\caption{The couplings $\sqrt{U_n}$ for even $N$ at $b=0$ ($N=40$, $\mu=0.8$), exact
from \eqref{eq:arcseven}: two arcs of the same radius $j+\tfrac12+\mu$, the odd bonds centred
at $y=-\mu$ and the even ones at $y=+\mu$. Sampled at the two bonds of a site, the two arcs differ by $2\mu\pm1$ times the
slope of the arc, which is where the superpotential $\mu\tan x$ comes from; the diagonal is the
constant staggered mass $\pm\tfrac\mu2$.}
\label{fig:even}
\end{figure}
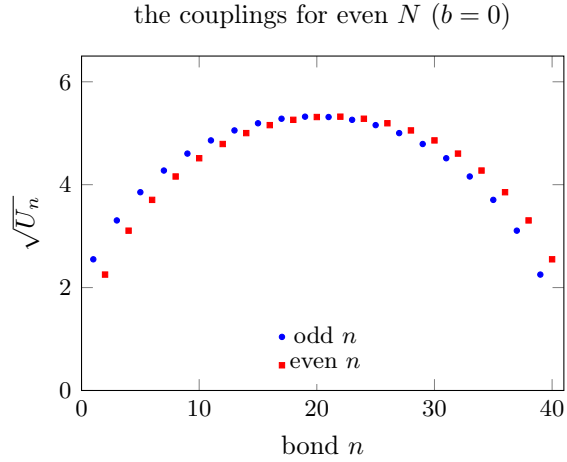

\paragraph{The case of even $N$.}
Let $N=2j$ with $j$ even (the case of odd $j$ is described at the end of the section). At
$\alpha=\half$ the recurrence coefficients of \cite[Eq.~(3.4)]{PBI} reduce, for every $b$, to
\begin{equation}
\label{eq:chaineven}
B_n=(-1)^n\,\frac{2\mu+b}4,\qquad
U_n=\begin{cases}
\dfrac{(n+2\mu)(N+1-n)}{16}\cdot\dfrac{n-j+b}{n-j}, & n\ \text{odd},\\[10pt]
\dfrac{n(N+1+2\mu-n)}{16}\cdot\dfrac{n-j-1-b}{n-j-1}, & n\ \text{even}.
\end{cases}
\end{equation}
Indeed, at $\alpha=\half$ the two exceptional entries of \cite[Eq.~(3.4)]{PBI},
$A_j=\tfrac12(1-\alpha)a$ and $C_j=\tfrac12\alpha a$, both equal $\tfrac a4$, which is the
value of the generic even--$n$ formulas $A_n=-\tfrac14(n-j-a)$, $C_n=\tfrac14(n-j+a)$ at
$n=j$; so \cite[Eq.~(3.4)]{PBI} holds without exception, with
$A_n=-\tfrac14(n+1)(n-j-b)/(n-j)$ and $C_n=\tfrac14(n-2j-1)(n-j+b)/(n-j)$ for odd $n$. Then
$B_n=\tfrac14(b-j-1+a)-A_n-C_n$ is $\tfrac14(b-j-1-a)=\tfrac14(2\mu+b)$ for even $n$, since
$-A_n-C_n=-\tfrac a2$, and $\tfrac14(b-j-1+a)+\tfrac12(j+1-b)=-\tfrac14(2\mu+b)$ for odd $n$,
since $-A_n-C_n=\tfrac{2(n-j)(j+1-b)}{4(n-j)}$; and $U_n=A_{n-1}C_n$ is, for odd $n$,
$-\tfrac1{16}(n-1-j-a)(n-2j-1)\tfrac{n-j+b}{n-j}$ with $n-1-j-a=n+2\mu$, and for even $n$
$-\tfrac1{16}\,n\,(n-j+a)\tfrac{n-1-j-b}{n-1-j}$ with $n-j+a=-(N+1+2\mu-n)$. Its eigenvalues
are the same numbers as for $N=2j+1$: the grid \eqref{eq:grid}, i.e.\ the towers
\eqref{eq:towers} minus $\tfrac14$, except that the odd sublattice stops at the level
$\ell=j-1$ and loses its outermost point. Explicitly, the spectrum of $J$ is
$\{\tfrac14(4n+2\mu+b)\}_{n\ge0}\cup\{-\tfrac14(4n+2\mu+b)\}_{n\ge1}$ (even sublattice)
together with $\{\pm\tfrac14(4n+2+2\mu-b)\}_{n\ge0}$ (odd sublattice), $n$ running as far as
the towers allow. Three things have changed. The grid now straddles $0$ rather than $c_0$:
$x_{\min}+x_{\max}=0$ exactly, the two--sided operator is $J$ itself and no shift is needed.
The diagonal of $J$ is a \emph{constant staggered mass} $\pm\tfrac14(2\mu+b)$, not zero: $J$
is not chiral, and the $\pm x$ pairing of its spectrum is broken by exactly one eigenvalue,
$\tfrac14(2\mu+b)$, the level $\ell=0$ of the even sublattice, whose would--be partner
$-\tfrac14(2\mu+b)$ is absent. And the two families of couplings, in terms of the bond
position $y=n-\tfrac12-\rho$, $\rho=\tfrac N2=j$, are
\begin{equation}
\label{eq:arcseven}
u(y)=\tfrac14\sqrt{M^2-(y+\mu)^2}\,\Big(1+\frac{b}{y+\frac12}\Big)^{1/2},\qquad
v(y)=\tfrac14\sqrt{M^2-(y-\mu)^2}\,\Big(1-\frac{b}{y-\frac12}\Big)^{1/2},
\end{equation}
with $M=j+\tfrac12+\mu$, since for a bond $n=y+\tfrac12+\rho$ one has
$(n+2\mu)(N+1-n)=M^2-(y+\mu)^2$ and $n(N+1+2\mu-n)=M^2-(y-\mu)^2$, while $n-j=y+\tfrac12$
and $n-j-1=y-\tfrac12$: two arcs of the \emph{same} radius $M$ (half a unit less than the
mean of the two radii $j+1+2\mu$ and $j+1$ of \eqref{eq:A:PR}), with centres displaced by
$\mp\mu$ (Fig.~\ref{fig:even}). The dimerization of \S\ref{sec:eqlimit} is thus realized
differently: not by two concentric arcs of different radii but by two congruent arcs shifted
against each other, and it is now accompanied by a mass.

\paragraph{The limit: matrix supersymmetry.}
The two--component recast \eqref{eq:spinor}--\eqref{eq:dirac-disc} applies verbatim (the last
site $2j$ is even, so $u_j=0$), with $\beta_n$ replaced by $B_n$, the operator being $J$
itself, and so does the bulk expansion \eqref{eq:A:expand}, with the arcs \eqref{eq:arcseven}
in place of \eqref{eq:A:PR}. With $g(y)=\tfrac14\sqrt{M^2-y^2}$, \eqref{eq:arcseven} reads
$u(y)=g(y+\mu)\big(1+\tfrac b{2y}+O(y^{-2})\big)$ and $v(y)=g(y-\mu)\big(1-\tfrac b{2y}+O(y^{-2})\big)$
for $|y|\gg1$, so that
\[
u-v=2\mu\,g'+\frac by\,g+O(N^{-1}),\qquad \tfrac12(u'+v')=g'+O(N^{-1}),\qquad u+v=2g+O(N^{-1}),
\]
and, by $y=\rho\sin x$, $g=\tfrac14\rho\cos x\,(1+O(N^{-1}))$, $g'=-\tfrac14\tan x$,
$\tfrac by\,g=\tfrac b4\cot x$ and $2g\,\partial_y=\tfrac12\partial_x$, all up to $O(N^{-1})$.
The difference with the odd case is that the two arcs $u$ and $v$ are now displaced rather
than of different radii: at an even site,
$u(y_0+\tfrac12)\chi(y_0+1)-v(y_0-\tfrac12)\chi(y_0-1)=\big[(2\mu+1)g'+\tfrac by g\big]\chi+2g\,\partial_y\chi$,
and at an odd site
$u(y_1-\tfrac12)\varphi(y_1-1)-v(y_1+\tfrac12)\varphi(y_1+1)=\big[(2\mu-1)g'+\tfrac by g\big]\varphi-2g\,\partial_y\varphi$.
With the masses $B_{2m}=m_b$, $B_{2m+1}=-m_b$, \eqref{eq:dirac-disc} contracts to
\begin{align*}
x_s\,\varphi&=m_b\varphi+\tfrac12\partial_x\chi-\Big(\frac{2\mu+1}4\tan x-\frac b4\cot x\Big)\chi,\\
x_s\,\chi&=-m_b\chi-\tfrac12\partial_x\varphi-\Big(\frac{2\mu-1}4\tan x-\frac b4\cot x\Big)\varphi ,
\end{align*}
where $x_s$ is the eigenvalue of $J$, and the normalization \eqref{eq:tilde},
$\tilde\varphi=(\cos x)^{1/2}\varphi$, $\tilde\chi=(\cos x)^{1/2}\chi$, absorbs
$\mp\tfrac14\tan x$ and leaves
\begin{equation}
\label{eq:diraceven}
x_s\,\tilde\varphi=m_b\,\tilde\varphi+\tfrac12\big(\partial_x-W_b\big)\tilde\chi,\qquad
x_s\,\tilde\chi=-m_b\,\tilde\chi-\tfrac12\big(\partial_x+W_b\big)\tilde\varphi,
\end{equation}
\begin{equation}
\label{eq:Wb}
W_b(x)=\mu\tan x-\frac b2\cot x,\qquad m_b=\frac{2\mu+b}4 .
\end{equation}
That is, with $\mathcal A_b=\tfrac12(\partial_x-W_b)$ and
$\mathcal A_b^\dagger=-\tfrac12(\partial_x+W_b)$,
\begin{equation}
\label{eq:Dirac}
J\ \longrightarrow\ \mathcal D_b=\begin{pmatrix} m_b&\mathcal A_b\\ \mathcal A_b^\dagger&-m_b\end{pmatrix},
\qquad
\mathcal D_b^2=\frac14\begin{pmatrix}\mathcal H_{\mu,\,b}&0\\0&\mathcal H_{\mu+1,\,b+2}\end{pmatrix},
\end{equation}
\begin{equation}
\label{eq:PTpair}
\mathcal H_{\mu,b}=-\partial_x^2+\frac{\mu(\mu-1)}{\cos^2x}+\frac{\frac b2(\frac b2-1)}{\sin^2x}\qquad\text{on }(-\tfrac\pi2,\tfrac\pi2),
\end{equation}
the squares following from
$W_b^2\pm W_b'=\mu(\mu\pm1)/\cos^2x+\tfrac b2(\tfrac b2\pm1)/\sin^2x-(\mu+\tfrac b2)^2$ and
$4m_b^2=(\mu+\tfrac b2)^2$, the constant being absorbed exactly. Like
$\mathcal D_\mu$ of \eqref{eq:A:Dmu}, the matrix operator \eqref{eq:Dirac} contains no
reflection. This is the ordinary, matrix supersymmetric quantum mechanics of \cite{CKS}:
$W_b=\mu\tan x-\tfrac b2\cot x$ is the superpotential of the trigonometric P\"oschl--Teller
potential, whose partners $\mathcal H_{\mu,b}$ and $\mathcal H_{\mu+1,b+2}$ form the
\emph{shape--invariant} pair of \cite[Table~4.1]{CKS} (the partner has the same form, with
$\mu$ and $\tfrac b2$ each shifted by one unit), and $J$ contracts to the supercharge $\mathcal A_b$
completed by the mass $m_b\sigma_z$, where $\sigma_z=\operatorname{diag}(1,-1)$ on the two
components anticommutes with the off--diagonal part; at $b=0$ the pair is
$\mu(\mu\mp1)/\cos^2x$. Where an even number of sites ($N$ odd) gave a first--order Dunkl
operator on \emph{one} function, an odd number of sites ($N$ even) gives a first--order matrix
operator on \emph{two}.

\paragraph{Eigenfunctions and spectrum.}
For $|b|<1$, the solutions of $\mathcal H_{\mu,b}\psi=E\psi$ that are square integrable on the
whole interval, with the wall behaviour $(\cos x)^{\mu}$, come in two families, distinguished
by their behaviour $|\sin x|^{b/2}$ or $\operatorname{sgn}(x)|\sin x|^{1-b/2}$ at the centre:
they are the P\"oschl--Teller functions
\begin{equation}
\label{eq:eigeven}
\begin{aligned}
\tilde\varphi&\propto(\cos x)^{\mu}\,|\sin x|^{b/2}\,\mathrm P_r^{(\frac b2-\frac12,\,\mu-\frac12)}(\cos2x),
&x_s&=\pm\tfrac14(4r+2\mu+b),\\[2pt]
\tilde\varphi&\propto(\cos x)^{\mu}\operatorname{sgn}(x)\,|\sin x|^{1-b/2}\,\mathrm P_r^{(\frac12-\frac b2,\,\mu-\frac12)}(\cos2x),
&x_s&=\pm\tfrac14(4r+2+2\mu-b),
\end{aligned}
\end{equation}
$r=0,1,\dots$, where $\mathrm P^{(\alpha',\beta')}_r$ denotes the classical Jacobi polynomial
\cite[\S9.8]{KLS} (upright, to distinguish it from the little $-1$ Jacobi polynomial
$P^{(A,B)}_\ell$), with the odd--site component $\tilde\chi=\mathcal A_b^\dagger\tilde\varphi/(x_s+m_b)$
and the eigenvalues $x_s$ of $\mathcal D_b$ indicated. At $b=0$ the even--site components are
the Gegenbauer functions $(\cos x)^\mu C_{2r}^{(\mu)}(\sin x)$ and $(\cos x)^\mu C_{2r+1}^{(\mu)}(\sin x)$
\cite[\S9.8.1]{KLS}, with the partners $(\cos x)^{\mu+1}C_{2r-1}^{(\mu+1)}(\sin x)$ and
$(\cos x)^{\mu+1}C_{2r}^{(\mu+1)}(\sin x)$ on the odd sites. The eigenvalues in
\eqref{eq:eigeven} are the exact spectrum of $J$ quoted above, provided the state $r=0$ of the
first family carries the sign $+$ only. This is the case, and it is decided at finite $N$. A
vector supported on the even sites alone, $p_{2m}=(-1)^m\varphi_m$, $p_{2m+1}=0$, satisfies
\eqref{eq:dirac-disc} if and only if $u_m\varphi_m=v_m\varphi_{m+1}$ for $m=0,\dots,j-1$ (the
odd sites) and $x_s=B_{2m}=\tfrac14(2\mu+b)$ (the even sites), the last even site $2j$
imposing nothing new since $u_j=0$. Hence $J$ has, exactly, the eigenvalue $\tfrac14(2\mu+b)$ ---
the level $\ell=0$ of the even sublattice --- with the eigenvector
$\varphi_m=\varphi_0\prod_{i<m}u_i/v_i$, i.e.\ by \eqref{eq:chaineven}
\begin{equation}
\label{eq:unpaired}
\varphi_m^2=\varphi_0^2\;
\frac{(\mu+\tfrac12)_m\,(-j)_m\,\big(\tfrac{1-j+b}2\big)_m}
{m!\,(\tfrac12-\mu-j)_m\,\big(\tfrac{1-j-b}2\big)_m}.
\end{equation}
In terms of the position $z=2m+1-j$ of the odd site between the bonds $u_m$ and $v_m$, the
ratio is $(u_m/v_m)^2=\frac{M^2-(z+\mu-\frac12)^2}{M^2-(z-\mu+\frac12)^2}\cdot\frac{z+b}{z-b}$,
so that, two units of $z$ separating consecutive even sites,
$\frac{d}{dz}\ln\varphi^2=-\frac{(2\mu-1)z}{M^2-z^2}+\frac bz+O(N^{-2})$ for $|z|\gg1$, whence
$\varphi\propto(M^2-z^2)^{(2\mu-1)/4}|z|^{b/2}\propto(\cos x)^{\mu-1/2}|\sin x|^{b/2}$ and
$\tilde\varphi\propto(\cos x)^{\mu}|\sin x|^{b/2}$. In the limit, this
$\tilde\varphi_0=(\cos x)^\mu|\sin x|^{b/2}$ is annihilated by $\mathcal A_b^\dagger$, so that
$(\tilde\varphi_0,0)$ is an eigenvector of $\mathcal D_b$ with eigenvalue $+m_b$ --- the
unpaired mode --- while the function $\tilde\chi\propto(\cos x)^{-\mu}|\sin x|^{-b/2}$
annihilated by $\mathcal A_b$, which would give the partner $(0,\tilde\chi)$ with eigenvalue
$-m_b$, is not the limit of any eigenvector: it is not square integrable for $\mu\ge\tfrac12$,
and for $\mu<\tfrac12$ it has the wall behaviour $(\cos x)^{-\mu}$ instead of the
$(\cos x)^{\mu+1}$ that all odd--site envelopes have in \eqref{eq:eigeven}.

The convergence of the eigenvectors to \eqref{eq:eigeven} follows from the polynomials as in
\S\ref{sec:polylimit}, and more simply, there being no Geronimus combination: the
amplitudes are $\varkappa_mF_m(x_s)/h_{2m}$ on the even sites and
$\varkappa'_m(x_s-m_b)G_m(x_s)/h_{2m+1}$ on the odd ones, with the series \eqref{eq:F43series} of
\cite[\S3.3]{PBI}. On the first sublattice, $x_s=\pm(m_b+n')$, the duality terminates
$F_m$ at $n'$ and $G_m$ at $n'-1$, and the confluence \eqref{eq:termlimit} gives
\begin{equation}
\label{eq:confeven}
\begin{aligned}
F_m(x_s)&\to{}_2F_1\big(-n',n'+2m_b;\mu+\tfrac12;\cos^2x\big)\propto \mathrm P^{(\frac b2-\frac12,\,\mu-\frac12)}_{n'}(\cos2x),\\
G_m(x_s)&\to{}_2F_1\big(1-n',n'+1+2m_b;\mu+\tfrac32;\cos^2x\big)\propto \mathrm P^{(\frac b2+\frac12,\,\mu+\frac12)}_{n'-1}(\cos2x);
\end{aligned}
\end{equation}
the even--site envelope $\varkappa_m/h_{2m}$ is now \emph{exactly} the unpaired mode
\eqref{eq:unpaired}, whose limit is also recovered from the closed form of the norm in
Appendix~\ref{app:C3}, and the odd--site envelope is
\[
\frac{\varkappa'_m}{h_{2m+1}}=\frac{\varkappa_m}{h_{2m}}\cdot\frac{\varkappa'_m}{\varkappa_mu_m},\qquad
\frac{\varkappa'_m}{\varkappa_mu_m}\to-\frac{2\sin x\cos x}{2\mu+1}
\]
by \eqref{eq:C:evenodd}. Restoring
$(\cos x)^{1/2}$, the even sites give the first family of \eqref{eq:eigeven} with $r=n'$, and
the odd sites give
\[
\tilde\chi\propto(\cos x)^{\mu+1}\sin x\,|\sin x|^{b/2}\,\mathrm P^{(\frac b2+\frac12,\,\mu+\frac12)}_{n'-1}(\cos2x),
\]
which is $\mathcal A_b^\dagger\tilde\varphi$ up to a constant --- the eigenfunction of
$\mathcal H_{\mu+1,b+2}$ with the exponent $\tfrac b2+1$ at the centre --- the two signs of $x_s$
differing only in the factor $x_s-m_b=n'$ or $-(2m_b+n')$, as $1/(x_s+m_b)$ requires. On the
second sublattice, $x_s=\pm(m'_b+n')$, the Whipple forms $F_m=\Pi_mF^W_m$, $G_m=\Pi'_mG^W_m$
terminate at $n'$ and give
\begin{equation}
\label{eq:confevenW}
\begin{aligned}
F^W_m(x_s)&\to{}_2F_1\big(-n',n'+2m'_b;\mu+\tfrac12;\cos^2x\big)\propto \mathrm P^{(\frac12-\frac b2,\,\mu-\frac12)}_{n'}(\cos2x),\\
G^W_m(x_s)&\to{}_2F_1\big(-n',n'+2m'_b;\mu+\tfrac32;\cos^2x\big)\propto \mathrm P^{(-\frac12-\frac b2,\,\mu+\frac12)}_{n'}(\cos2x),
\end{aligned}
\end{equation}
while the factors $\Pi_m\propto|\sin x|^{1-b}$ and $\Pi'_m\propto|\sin x|^{1-b}/\sin^2x$ change
the exponents at the centre to $1-\tfrac b2$ and $-\tfrac b2$: the second family of
\eqref{eq:eigeven} with $r=n'$, and its partner
\[
\tilde\chi\propto(\cos x)^{\mu+1}|\sin x|^{-b/2}\,\mathrm P^{(-\frac12-\frac b2,\,\mu+\frac12)}_{n'}(\cos2x),
\]
the second Frobenius solution of $\mathcal H_{\mu+1,b+2}$ at the centre. All this is on the left half
of the lattice, where the polynomial representations hold; the mirror, to which we now turn,
carries it to the right half and supplies the parities --- in particular the
$\operatorname{sgn}(x)$ of the second family.

\paragraph{Mirror, channels and the reflection in the Hamiltonian.}
The site reversal $n\mapsto N-n$ now preserves the parity of the sites, and by the staggering
\eqref{eq:spinor} it acts on the two--component field as $\mathcal M'=\sigma_z\R$,
$(\tilde\varphi,\tilde\chi)\mapsto(\R\tilde\varphi,-\R\tilde\chi)$ --- not the exchange
$\mathcal M$ of \eqref{eq:Mexch} --- which commutes with
$\mathcal D_b$ because $W_b$ is odd. The two channels are its eigenspaces: on the channel
$\epsilon=+1$ the even--site envelope is even and the odd--site envelope odd, on the channel
$\epsilon=-1$ the reverse --- the two families of \eqref{eq:eigeven}, which carry the two
Frobenius exponents $\tfrac b2$ and $1-\tfrac b2$ of $\mathcal H_{\mu,b}$ at the centre
(both square--integrable for $|b|<1$, so that the differential expression leaves the
behaviour there open), graded by the mirror as in the two companions. Now put the two components
together, $\psi=\tilde\varphi+\tilde\chi$, a function whose even and odd parts are the two
envelopes. Using $\R\partial_x=-\partial_x\R$ and the oddness of $W_b$, the system
\eqref{eq:diraceven} becomes $x_s\,\psi=-\tfrac12\big[(\partial_x-2m_b)\R+W_b\big]\psi$ on
the channel with $\tilde\varphi$ even and $x_s\,\psi=\tfrac12\big[(\partial_x-2m_b)\R-W_b\big]\psi$
on the other, that is, on the channel $\epsilon=\pm1$,
\begin{equation}
\label{eq:Dunkleven}
J\ \longrightarrow\ \mp\tfrac1{\sqrt2}\,Q[\mathsf U,\pm\mathsf V],\qquad
\mathsf U=-\frac{2\mu+b}2,\quad \mathsf V=W_b ,
\end{equation}
with $Q[\mathsf U,\mathsf V]$ the supercharge with reflections \eqref{eq:postQ} of \cite{PVZ}
--- but a different member of the family: $\mathsf U$ is constant and $\mathsf V$ is the
P\"oschl--Teller superpotential, whereas for odd $N$ one had $\mathsf U=-\mu/\cos x$ and
$\mathsf V=\mp b/(2\sin x)$. Its square is, on the channel $\epsilon=\pm1$,
\begin{equation}
\label{eq:DunklPT}
J^2\ \longrightarrow\ \tfrac12Q[\mathsf U,\pm\mathsf V]^2=\frac14\Big[-\partial_x^2+\frac{\mu(\mu\mp\R)}{\cos^2x}
+\frac{\frac b2(\frac b2\mp\R)}{\sin^2x}\Big],
\end{equation}
the Dunkl (reflection--graded) P\"oschl--Teller Hamiltonian, whose two parity sectors are the
two partners of \eqref{eq:Dirac}, in the order fixed by the channel. So the parity of $N$ decides where the reflection lives.
For odd $N$ it is in the supercharge and, at $b=0$, absent from the Hamiltonian; for even $N$
it is in the Hamiltonian, as the Dunkl potential $\mu(\mu-\R)/\cos^2x$ --- the trigonometric
analogue of the Wigner--Dunkl oscillator that the para--Krawtchouk limit turned out not to be
\cite{PKcont}, and the two wall exponents $\mu$, $\mu+1$ of the two sectors differ by
exactly one, as a Dunkl potential requires. The two limits of $J$ are isospectral --- the
eigenvalues of $J$ are given by the same formulas for both parities --- but they are different
operators: $\Qs$ acts on one function and its square is a scalar Scarf~I operator,
$\mathcal D_b$ acts on two and its square is a diagonal P\"oschl--Teller pair. The row ``role
of $\R$'' of the table of \S\ref{sec:compare} thus has two entries for the para--Bannai--Ito
Hamiltonian: for odd $N$ the reflection enters the supercharge (and, for $b\neq0$, the
Hamiltonian as well, \S\ref{sec:reflection}), for even $N$ the Hamiltonian.

\paragraph{Even $N$, odd $j$.}
For $N=2j$ with $j$ odd the coefficients of \cite[App.~A]{PBI} give, at $\alpha=\half$, the
same $B_n$ as in \eqref{eq:chaineven} with $b\to-b$ and the same $U_n$ with the $b$--dependent
factors of the two parities exchanged, $\tfrac{n-j-1-b}{n-j-1}$ on odd and $\tfrac{n-j+b}{n-j}$
on even $n$, which in the bulk is again $b\to-b$: $J$ contracts to $\mathcal D_{-b}$, with the
eigenfunctions and eigenvalues \eqref{eq:eigeven} with $b\to-b$, and the spectrum of $J$ is
\eqref{eq:towersodd} minus $\tfrac14$, with the odd sublattice stopping at $\ell=j-1$. The
staggering \eqref{eq:spinor} acquires the sign $(-1)^j$ under the site reversal, which
therefore acts on the two--component field as $-\mathcal M'$: on the channel $\epsilon=+1$ the
even--site envelope is now odd and the odd--site envelope even, and the reverse on
$\epsilon=-1$. In particular the unpaired mode, of eigenvalue $\tfrac14(2\mu-b)$ and even on
the even sites, has $\epsilon=-1$ and is the level $\ell=0$ of the odd sublattice. The largest
eigenvalue is still $x_0$, on the even sublattice, where $\epsilon=+1$.

\section{Algebraic structure: the bispectral pair and its Dunkl realization}
\label{sec:algebra}

The para--Bannai--Ito polynomials are \emph{bispectral}: besides the recurrence \eqref{eq:rec} they
satisfy a Dunkl--difference equation in the spectral variable, which we denote by $t$ in this
section, $L\,P_n(t)=\Lambda_n P_n(t)$, with \cite[\S4.2]{PBI}
\begin{equation}
\label{eq:Ldunkl}
L=F_L(t)\,(I-\R_t)+G_L(t)\,\big(T^+\R_t-I\big),\qquad \Lambda_{2m}=m,\ \ \Lambda_{2m+1}=j-m,
\end{equation}
where $\R_tf(t)=f(-t)$, $T^+f(t)=f(t+1)$ and $F_L,G_L$ are rational functions of $t$ given in
\cite{PBI}. In the eigenvector picture, where $\ket{s}$ has the components $\sqrt{w_s}\,p_n(x_s)$,
$J$ acts as multiplication by $x_s$ on the eigenvector and $L$ as multiplication by $\Lambda_n$
on the $n$-th component: $J$ is diagonal in the grid, $L$ in the degree. The pair $(J,L)$
generates a representation of the \emph{Bannai--Ito algebra} $\bialg$, the $q=-1$ Askey--Wilson
algebra $\awalg$ \cite{BI,TVZ} (gothic letters denote the algebras named after the polynomial
families, the subscript their rank), surveyed with its realizations and applications in
\cite{DGTVZ}; this representation is reducible, the direct sum of two Bannai--Ito modules carried
by the two sublattices \cite{BCLMNV}, with structure constants that depend on the size $j$ of the
model. We shall not contract these structure constants; we determine instead what the two
bispectral operators become in the limit, and observe that they realize the anticommutator
algebra attached to the little $-1$ Jacobi polynomials in \cite[\S6]{VZm1}.

The limit of $J$ is the content of \S\ref{sec:reflection}: on each channel, $J+\tfrac14$ tends to
$\mp\tfrac1{\sqrt2}Q_{\pm b,2\mu}$ \eqref{eq:limitb}, a first--order Dunkl operator. The limit of
$L$ is elementary and exact: by \eqref{eq:x}, the site $n$ sits at the position
$\xi_n$ with $\sin\xi_n=n/\rho-1$, $\rho=N/2$, so that
$\Lambda_{2m}=m=\tfrac\rho2(1+\sin\xi_{2m})$ and
$\Lambda_{2m+1}=j-m=\tfrac\rho2(1-\sin\xi_{2m+1})$, i.e.
\begin{equation}
\label{eq:Llimit}
1-\frac2\rho\,L=-(-1)^n\sin\xi_n\quad\text{on the $n$-th component}:
\end{equation}
on the two--component field, $1-\tfrac2\rho L$ acts as multiplication by $-\sin x$ on the
even--site envelope $\varphi$ and by $+\sin x$ on the odd--site envelope $\chi$, exactly at
finite $N$. The roles of the two bispectral operators are thus \emph{exchanged} in the limit:
the recurrence operator $J$, diagonal in the grid, becomes a differential--difference (Dunkl)
operator, and the Dunkl--difference operator $L$, diagonal in the degree, becomes a
multiplication operator in the position. This is the duality of \S\ref{sec:polylimit} seen on
the operators.

In the limit both operators act on the eigenfunctions $\Psi_{\ell;A,B}(x)$, $A=\pm b$, $B=2\mu$,
which by \eqref{eq:m1wave} are the little $-1$ Jacobi polynomials $P^{(A,B)}_\ell(\zeta)$ in the
variable $\zeta=\sin x$ times the square root of their weight. Stripped of this gauge factor,
the supercharge $Q_{A,B}$ becomes an operator on polynomials in $\zeta$ with the same
eigenvalues, $\tfrac1{2\sqrt2}(-1)^{\ell+1}(2\ell+A+B+1)$, and the Dunkl operator $L_0$ of
\eqref{eq:m1eig} has, on the same polynomials, the eigenvalues $\omega_\ell=-2\ell$ ($\ell$
even), $2(\ell+A+B+1)$ ($\ell$ odd). Comparing,
\begin{equation}
\label{eq:awm1}
X=\tfrac{A+B+1}2-\tfrac12L_0,\qquad Y=\zeta,\qquad Z=(1-\zeta)\R
\end{equation}
is the triple of \cite[Eq.~(6.7)]{VZm1} written in our convention for $L_0$: $X$ has the
eigenvalues $(-1)^{\ell}\tfrac12(2\ell+A+B+1)$ on $P^{(A,B)}_\ell$, so that $X/\sqrt2$ is, up to
the gauge factor and the change of variable $\zeta=\sin x$, the operator $-Q_{A,B}$ --- the
limit of $\pm\sqrt2\,(J+\tfrac14)$ on the channels $A=\pm b$, by \eqref{eq:limitb} --- while $1-\tfrac2\rho L$ acts on the
even--site envelope as multiplication by $-\sin x=\sin(-x)$, which is precisely the variable
$\zeta$ of the limit $\Psi_\ell(-x)$ of that envelope \eqref{eq:evenlimit}: it is the operator
$Y$. A direct computation on a generic function of $\zeta$ gives the \emph{anticommutator}
relations
\begin{equation}
\label{eq:awrel}
\{X,Y\}=Z+A,\qquad \{Y,Z\}=0,\qquad \{Z,X\}=Y+B,\qquad Y^2+Z^2=I ,
\end{equation}
the last one being the Casimir element of the algebra in this realization \cite[\S6]{VZm1}.
These are the $q=-1$ Askey--Wilson relations of \cite[\S6]{VZm1}, a degenerate form of the
Bannai--Ito algebra in which the anticommutator $\{Y,Z\}$ has no linear term; the reflection
enters the generators themselves through $Z=(1-\zeta)\R$. The Hamiltonian $H=\Qs^2$ and the
supercharge $\Qs$ are built from these generators, and the ladder representation of
\eqref{eq:awrel} on the polynomial basis reproduces the little $-1$ Jacobi eigenvalues
\eqref{eq:spec}, \eqref{eq:towers} and eigenfunctions \cite{VZm1}. The same algebra arises as the Racah algebra
of $sl_{-1}(2)$ --- the paraboson algebra supplemented by the reflection, i.e.\ the odd
generators of $osp(1|2)$ together with an involution --- and as a symmetry algebra of the
Dunkl--Dirac operator on the sphere, a first--order operator whose square gives the Dunkl
Laplacian \cite{DGTVZ}. The chiral--supersymmetric reading of such square roots is raised
there as an open question; the para--Bannai--Ito limit supplies an instance in which the
first--order operator is an honest supercharge, $H=\Qs^2$, on a bounded interval.

At the algebraic level the trilogy thus descends one line of the $q$--Askey scheme, in step
with the confluence of the polynomials: the bispectral pair of the para--Krawtchouk
oscillator, which generates the Hahn algebra $\halg$, becomes a pair of generators of the
Lie algebra $\su$ of the singular oscillator \cite{PKcont}; that of the para--Racah
Hamiltonian, which generates the Racah algebra $\ralg$, becomes a pair of generators of the
quadratic Jacobi algebra $\jalg$ \cite{PRcont}; and that of the para--Bannai--Ito
Hamiltonian, which generates the Bannai--Ito ($q=-1$ Askey--Wilson) algebra $\bialg$, becomes
the Dunkl realization of the anticommutator relations \eqref{eq:awrel} --- the fingerprint of
the reflection that, here alone, has entered the generators themselves.

\section{Comparison: the three Hamiltonians}
\label{sec:compare}

The three continuum limits now form a complete picture, organized by the arithmetic of the grid.

\begin{center}
\footnotesize
\renewcommand{\arraystretch}{1.3}
\setlength{\tabcolsep}{2.4pt}
\begin{tabular}{@{}llll@{}}
\toprule
 & para--Krawtchouk \cite{PKcont} & para--Racah \cite{PRcont} & para--Bannai--Ito (this paper)\\
\midrule
bi--lattice & linear $\{2s,2s{+}\gamma'\}$ & quadratic $\{(s{+}a')^2,(s{+}c')^2\}$ & linear, \emph{two--sided}\\
$q$--Askey scheme & $q\to1$ (Hahn) & $q\to1$ (Racah) & $q\to-1$ (Bannai--Ito)\\
expansion point & band edge & band edge & band \emph{centre} (Dirac)\\
order of limit & 2nd (Schr\"odinger) & 2nd (Schr\"odinger) & 1st (supercharge), $H{=}Q^2$\\
confluence & Hahn $\to$ Laguerre & Racah $\to$ Jacobi & $q{\to}{-}1$: little $-1$ Jacobi\\
domain & line $\mathbb{R}$ & interval $(0,\pi)$ & interval $(-\tfrac\pi2,\tfrac\pi2)$\\
limit Hamiltonian & isotonic osc.\ & trig.\ P\"oschl--Teller & gen.\ P\"oschl--Teller (Scarf~I)\\
eigenfunctions & generalized Hermite & weighted Jacobi & little $-1$ Jacobi, $\Psi_{\ell;\pm b,2\mu}$\\
spectrum of $H$ & linear, $4t'{+}2{\mp}\gamma'$ & quadratic, $(t'{+}a')^2,(t'{+}c')^2$ & quadratic, $\tfrac1{16}(2\ell{+}1{+}2\mu{\pm}b)^2$\\
bispectral pair, finite $N$ & Hahn algebra $\halg$ & Racah algebra $\ralg$ & Bannai--Ito algebra $\bialg$\\
its limit generates & $\su$ (Lie) & $\jalg$ (quadratic) & anticommutator algebra \eqref{eq:awrel}\\
role of $\R$ & grades the two extensions & grades two channels & $\sqrt H$ (odd $N$), $H$ (even $N$)\\
\bottomrule
\end{tabular}
\end{center}

The unifying statement is that a continuous offset between two interleaved lattices produces,
in the limit, a reflection that combines two channels into one system; what the grid's
\emph{shape} dictates is the operator those channels live on. A linear grid bounded below leads
to the singular oscillator on the line (Laguerre functions, linear spectrum); a quadratic grid
to the P\"oschl--Teller box with singular walls (Jacobi functions, quadratic spectrum); a $q=-1$,
two--sided grid has no band edge at all, and the limit is a first--order Dunkl operator --- the
reflection enters the dynamics and the Hamiltonian is its square (\S\ref{sec:reflection}),
while the algebra generated by the limit of the bispectral pair descends from a Lie, through
a quadratic, to a $q=-1$ anticommutator algebra (\S\ref{sec:algebra}). (In the table,
$t'$ is the level label of the companions, $\gamma'$ the para--Krawtchouk offset and $a',c'$
the para--Racah parameters, primed to avoid the symbols of this paper.)

\section{Perfect state transfer, fractional revival and the persymmetric point}
\label{sec:FR}

This section settles two questions: when does the para--Bannai--Ito Hamiltonian transfer or
revive, and what of this survives the limit.

\paragraph{One--excitation dynamics, state transfer and revival.}
The matrix $J$ generates a dynamics of physical interest. An $XX$ spin chain of $N+1$ sites
with nearest--neighbour couplings $\sqrt{U_n}$ and local magnetic fields $B_n$ conserves the number of
spins up, and on its one--excitation subspace --- spanned by the states $\ket{e_n}$ in which the
single spin up sits at site $n$ --- its Hamiltonian acts precisely as $J$ \cite{VZ,paraRacahChain}.
We shall not introduce the spin chain itself; all that matters here is that the transport of an
excitation along it is the evolution $e^{-\ii \tau J}$, $\tau\ge0$, on $\mathbb C^{N+1}$, whose generator is the
discrete Hamiltonian of this paper. Two properties of this evolution are of interest \cite{GVZ}.
\emph{Perfect state transfer} at time $T$ means that an excitation launched at one end arrives
in full at the other, $e^{-\ii TJ}\ket{e_0}=e^{\ii\phi}\ket{e_N}$; \emph{fractional revival} at
time $T$ means that it reappears as a coherent superposition localized at the two ends,
\begin{equation}
\label{eq:FRdef}
e^{-\ii TJ}\ket{e_0}=\xi\,\ket{e_0}+\zeta\,\ket{e_N},\qquad |\xi|^2+|\zeta|^2=1,
\end{equation}
with probability $|\xi|^2$ of being found back at the origin and $|\zeta|^2$ of having been
transferred, perfect transfer being the case $\xi=0$. Both are read off the spectral data of
$J$, as follows. At the persymmetric point $\alpha=\half$, by \eqref{eq:eigvec},
$\ket{e_0}=\sum_s\sqrt{w_s}\,\ket{s}$ ($p_0=1$), and by \eqref{eq:eps},
$\ket{e_N}=\R_c\,\ket{e_0}=\sum_s\epsilon_s\sqrt{w_s}\,\ket{s}$, so that \eqref{eq:FRdef}
holds if and only if
\begin{equation}
\label{eq:FRphases}
e^{-\ii Tx_s}=\xi+\zeta\,\epsilon_s\qquad\text{for every }s,
\end{equation}
that is, if and only if the phases $e^{-\ii Tx_s}$ take a single value $e^{\ii\phi_+}$ on
the eigenvalues with $\epsilon_s=+1$ and a single value $e^{\ii\phi_-}$ on those with
$\epsilon_s=-1$. Then $\xi=\tfrac12(e^{\ii\phi_+}+e^{\ii\phi_-})$,
$\zeta=\tfrac12(e^{\ii\phi_+}-e^{\ii\phi_-})$, and since $I=\sum_s\ket{s}\bra{s}$ and
$\R_c=\sum_s\epsilon_s\ket{s}\bra{s}$, \eqref{eq:FRphases} is the operator identity
\begin{equation}
\label{eq:FRop}
e^{-\ii TJ}=e^{\ii\phi}\big(\cos\vartheta\,I+\ii\sin\vartheta\,\R_c\big)=e^{\ii\phi}\,e^{\ii\vartheta \R_c},
\qquad \phi=\tfrac12(\phi_++\phi_-),\quad \vartheta=\tfrac12(\phi_+-\phi_-).
\end{equation}
At time $T$ every state, not only $\ket{e_0}$, becomes the superposition of itself and its
mirror image, with the amplitudes $\cos\vartheta$ in place and $\ii\sin\vartheta$ on the image
(probabilities $\cos^2\vartheta$ and $\sin^2\vartheta$ when the state is orthogonal to its
mirror image, as $\ket{e_0}$ is). Perfect state transfer is $\vartheta\equiv\tfrac\pi2\pmod\pi$,
and at the multiples $kT$ of the time the angle is $k\vartheta$, $\R_c$ being an involution.
Since $\epsilon=+1$ on the even sublattice and $-1$ on the odd one \eqref{eq:eps}, the
criterion reads: \emph{$e^{-\ii Tx_s}$ must be constant on each of the two sublattices.}

\paragraph{Transfer on the two--sided grid.}
On the two--sided grid each sublattice consists of \emph{two} arms of unit spacing, one on
each side of $c_0$: in terms of $q=x_s-c_0$, \eqref{eq:grid} gives
\begin{equation}
\label{eq:arms}
\{x_{2s}\}\!:\ q\in\Big\{\tfrac{1+2\mu}{4}+k\Big\}\cup\Big\{-\tfrac{3+2\mu+2b}{4}-k\Big\},\qquad
\{x_{2s+1}\}\!:\ q\in\Big\{-\tfrac{1+2\mu}{4}-k\Big\}\cup\Big\{\tfrac{3+2\mu-2b}{4}+k\Big\},
\end{equation}
$k=0,1,\dots$, which is \eqref{eq:towers} displaced by $-\tfrac b4$. Unit spacing forces $T=2\pi \nu$ with $\nu$ a positive
integer, and the two arms of a sublattice are then in phase precisely when
\begin{equation}
\label{eq:PSTcond}
\nu(2\mu+b)\in2\mathbb Z,\qquad \nu(2\mu-b)\in2\mathbb Z .
\end{equation}
Under \eqref{eq:PSTcond} the two phases are $e^{\ii\phi_\pm}=e^{\mp\ii\pi \nu(1+2\mu)/2}$ and
\eqref{eq:FRop} reads
\begin{equation}
\label{eq:PSTid}
e^{-2\pi\ii \nu\,Q}=\cos\vartheta\,I+\ii\sin\vartheta\,\R_c ,\qquad
\vartheta=-\tfrac{\pi \nu}{2}(1+2\mu),
\end{equation}
an exact identity at finite $N$ (for $J$ itself the global phase $e^{-2\pi\ii \nu c_0}$ is restored).
But \eqref{eq:PSTcond} implies $2\nu\mu\in\mathbb Z$, hence
$\vartheta\in\tfrac\pi2\mathbb Z$: the right side is $\pm I$ when $\nu(1+2\mu)$ is even and $\pm\ii\R_c$
when it is odd. The persymmetric para--Bannai--Ito Hamiltonian therefore either returns to itself or
transfers states \emph{perfectly}; it never revives fractionally. This is a feature of the Bannai--Ito
grid. On the companions' bi--lattices the offset between the two sublattices leaves the revival
angle free (para--Krawtchouk, through the offset of its two sublattices) or quantized
but generically fractional (para--Racah \cite{paraRacahChain,PRcont}); here each sublattice
straddles the centre, and the coherence of its two arms quantizes the angle to multiples of
$\pi/2$. (For instance
$\mu=\tfrac45$, $b=\tfrac25$, $\nu=5$ satisfy \eqref{eq:PSTcond} with $\nu(1+2\mu)=13$ odd, and
\eqref{eq:PSTid} with the global phase restored gives $e^{-10\pi\ii J}=-\R_c$: perfect transfer
at $T=10\pi$.)

\paragraph{Revival through the isospectral deformation.}
Fractional revival in this model is entirely the work of the isospectral deformation. The
para--Bannai--Ito Hamiltonian is persymmetric --- it commutes with the site reversal $\R_c$
\cite{GTVZ} --- exactly at $\alpha=\half$ (\S\ref{sec:chain}). The construction of
\cite[\S4]{GTVZ} deforms a persymmetric matrix by conjugation with the orthogonal
involution $\mathcal K=\sin\theta\,\Sigma+\cos\theta\,\R_c$, where $\Sigma=+1$ on the sites
$0,\dots,j$ and $-1$ on the sites $j+1,\dots,N$, as in \cite[Eq.~(4.1)]{GTVZ}: $\tilde J=\mathcal KJ\mathcal K$ has the same spectrum, is no longer persymmetric, and differs from $J$ in three
entries only --- the central coupling, multiplied by $\cos2\theta$, and the two central diagonal
entries, shifted by $\pm\sqrt{U_{j+1}}\,\sin2\theta$ \cite[Eq.~(4.5)]{GTVZ}. Comparing with
\eqref{eq:AC}, where $\sqrt{U_{j+1}}=\tfrac{|a|}{4}\cdot2\sqrt{\alpha(1-\alpha)}$ and
$B_j,B_{j+1}=B_j(\half)\pm\tfrac{|a|}{4}(2\alpha-1)$, shows that the para--Bannai--Ito family is
exactly this deformation,
\begin{equation}
\label{eq:GTVZ}
J(\alpha)=\mathcal K\,J(\tfrac12)\,\mathcal K,\qquad \sin2\theta=2\alpha-1 .
\end{equation}
Two consequences follow. The weights are rescaled sublattice by
sublattice, $\tilde w_s=(1\pm\sin2\theta)\,w_s$ \cite[Prop.~4.1]{GTVZ}, i.e.\ by $2\alpha$ on
$\{x_{2s}\}$ and by $2(1-\alpha)$ on $\{x_{2s+1}\}$, whose total weights become $\alpha$ and $1-\alpha$
\cite{PBI}: the grid does not move, only the measure is redistributed between the two channels. And at
a transfer point --- \eqref{eq:PSTcond} with $\nu(1+2\mu)$ odd --- conjugating \eqref{eq:PSTid}
by $\mathcal K$ gives $e^{-2\pi\ii \nu\,Q(\alpha)}=\pm\ii\,\mathcal K\R_c\mathcal K$ with
$\mathcal K\R_c\mathcal K\ket{e_0}=\sin2\theta\ket{e_0}+\cos2\theta\ket{e_N}$:
the excitation is split between the two ends with probabilities
\begin{equation}
\label{eq:FRamp}
(2\alpha-1)^2\ \text{ at site }0,\qquad 4\alpha(1-\alpha)\ \text{ at site }N ,
\end{equation}
exact at finite $N$. The deformation parameter of the
para--Bannai--Ito polynomials is thus the fractional--revival parameter of the model, in the sense of
\cite{GVZ,paraRacahChain}.

\paragraph{In the limit.}
In the continuum the criterion is applied to the limiting operator itself. At $b=0$, where the
operator is explicit, the supercharge $\Qs$ of one channel --- the odd sublattice, $\epsilon=-1$
(\S\ref{sec:eqlimit}) --- has the eigenvalues $(-1)^{\ell+1}\tfrac12(\ell+\mu+\tfrac12)$,
$\ell=0,1,2,\dots$ \eqref{eq:spec}, so that at $T=2\pi\nu$
\[
e^{-2\pi\ii \nu\Qs}\Psi_\ell=e^{(-1)^\ell\ii\pi \nu(2\ell+1+2\mu)/2}\,\Psi_\ell ,
\]
which is $e^{-\ii\vartheta}\Psi_\ell$ for even $\ell$ and $(-1)^\nu e^{\ii\vartheta}\Psi_\ell$ for odd $\ell$;
under \eqref{eq:PSTcond} at $b=0$, i.e.\ $\nu\mu\in\mathbb Z$, one has $(-1)^\nu=e^{-2\ii\vartheta}$
and the two agree. Hence $e^{-2\pi\ii \nu\Qs}=e^{-\ii\vartheta}I$ on this channel and, the other
channel carrying $-\Qs$ (\S\ref{sec:reflection}), $e^{\ii\vartheta}I$ on the even one: on each
channel the evolution at $T$ is a pure phase --- $\pm1$ at a return, $\pm\ii$ at a transfer --- and
the two channels acquire opposite phases. On the two--component field $(\varphi,\chi)$, whose
channels are the eigenspaces $\chi=\epsilon\R\varphi$ of the mirror $\mathcal M$ of
\eqref{eq:Mexch}, this is
$e^{-2\pi\ii \nu\,\mathcal D_\mu}=\cos\vartheta\,I+\ii\sin\vartheta\,\mathcal M$ for the limiting
two--component operator $\mathcal D_\mu$ of \eqref{eq:A:Dmu} (which is $\mp\Qs$ on the two
channels), the identity
\eqref{eq:PSTid} with $\R_c$ replaced by its continuum image $\mathcal M$; the same $T$ and
$\vartheta$ serve, both being fixed by $\mu$ alone, and the deformation acts on the channels as on
the sublattices. Read on the two--component field, this is revival in the sense of wave--packet dynamics
\cite{AP,Robinett}: at $T=2\pi \nu$ every state becomes a combination of itself and its mirror image
$(\varphi,\chi)\mapsto(\R\chi,\R\varphi)$ --- a packet near one wall in one component reappears near
the other wall in the other component --- with all the weight on the image at a transfer point;
the deformation splits the weight of a packet supported in one half, $(2\alpha-1)^2$ in place and
$4\alpha(1-\alpha)$ on the image, and only then is the revival fractional. The mirror $\R_c$ thus
survives as the channel sign $\epsilon$, \emph{not} as the position reflection $\R$ inside $\Qs$,
with which $H$ does not commute. The deformation survives as the
reweighting of the two channels by $2\alpha$ and $2(1-\alpha)$: the eigenvectors of
$J(\alpha)$ are $\mathcal K\ket{s}$, i.e.\ the eigenvectors of $J(\half)$ multiplied by
$\sin\theta+\epsilon_s\cos\theta$ on one half of the lattice and by
$-\sin\theta+\epsilon_s\cos\theta$ on the other \cite[\S4]{GTVZ}, the spectrum being fixed,
and the same holds for the limiting eigenfunctions on the two halves of the interval. The second
grid parameter $b$ enters only the arithmetic condition \eqref{eq:PSTcond} and the spectral
symmetry: at $b=0$ the grid is symmetric about $c_0$ and the $\pm q$ pairing is exact, for
$b\neq0$ the two towers \eqref{eq:towers} are displaced by $\pm b/4$ (even and odd sublattice
respectively) and the pairing is lost.
The doubly clean point is $\alpha=\half$, $b=0$: persymmetric and spectrally symmetric.

\paragraph{The other parity classes.}
For $N$ odd and $j$ odd, where the signs of the two supercharges are reversed
(\S\ref{sec:even}), the sign of $\vartheta$ in \eqref{eq:PSTid} is reversed, which changes the
right side by at most a global sign, $2\vartheta$ being a multiple of $\pi$ under
\eqref{eq:PSTcond}. For $N$ even the transfer identity \eqref{eq:PSTid} holds with the same
$T=2\pi\nu$, the same condition \eqref{eq:PSTcond} and the same angle modulo $\pi$ (a global
phase apart, $J$ being no longer shifted), the two mirror classes being the sublattices as
before, since the spectrum is the same. The isospectral deformation of
\cite[Eq.~(4.6)]{GTVZ} for an odd number of sites acts on the two \emph{central couplings},
multiplied by $\cos\theta\pm\sin\theta$, and leaves the diagonal alone; in
\cite[Eq.~(3.4)]{PBI} the $\alpha$--dependence sits exactly there,
$\sqrt{U_j}\propto\sqrt{2\alpha}$ and $\sqrt{U_{j+1}}\propto\sqrt{2(1-\alpha)}$, so that
\eqref{eq:GTVZ} holds with the same $\sin2\theta=2\alpha-1$ and the involution of
\cite[Eq.~(4.2)]{GTVZ}, which fixes the central site. The weights are again rescaled by
$2\alpha$ and $2(1-\alpha)$ on the two sublattices \cite[\S3.4]{PBI}, and at a transfer point
the excitation is split with the probabilities \eqref{eq:FRamp}. In the continuum the identity
reads $e^{-2\pi\ii\nu\,\mathcal D_b}=\cos\vartheta\,I\pm\ii\sin\vartheta\,\mathcal M'$ on the
two--component field, with $\mathcal M'=\sigma_z\R$ in place of the exchange $\mathcal M$: the
mirror survives as the grading of the channels for both parities, and it is the operator it
grades that changes.

\section{Conclusion}
\label{sec:concl}

The continuum limit of the para--Bannai--Ito Hamiltonian --- the operator defined the tridiagonal
Jacobi matrix of the eponymous polynomials --- is a supersymmetric system with reflections. Because the bi--lattice is
\emph{two--sided}, the natural object is the shifted operator $Q=J-c_0$, $c_0=(b-1)/4$, whose
spectrum straddles zero; it is a first--order \emph{Dunkl
supercharge}, the square root of a Schr\"odinger operator. Recasting the recurrence as a
two--component Dirac system and contracting it about the band centre --- a Dirac point, forced by the
chirality of $Q$ --- we obtained
\[
\Qs=\tfrac12\Big(\partial_x-\frac{\mu}{\cos x}\Big)\R,\qquad
H=\Qs^2=-\tfrac14\,\partial_x^2+\frac14\,\frac{\mu(\mu-\sin x)}{\cos^2x},
\]
the generalized P\"oschl--Teller (Scarf~I) Hamiltonian --- a differential operator without
reflection, whose square root $\Qs$ is a Dunkl operator, as in the supersymmetric quantum
mechanics with reflections of \cite{PVZ} --- with eigenfunctions the little $-1$ Jacobi
functions and spectrum $\tfrac14(\ell+\mu+\tfrac12)^2$.
 The bi--lattice supplies two such channels, combined by the
mirror $\R_c$ into one system.
For $b\neq0$ the operator $J+\tfrac14$ contracts on the two channels to the supercharges
$\pm\tfrac1{\sqrt2}Q_{\mp b,2\mu}$ of \cite{PVZ}, whose squares are the two--parameter extended
Scarf~I Hamiltonians with reflection, with little $-1$ Jacobi eigenfunctions and the exact spectrum
\eqref{eq:towers}, the Bannai--Ito bi--lattice itself. The parameter $\alpha$ governs persymmetry: at
$\alpha=\tfrac12$ the model transfers states perfectly at the time $T=2\pi\nu$ when
$\nu(2\mu\pm b)\in2\mathbb Z$ and $\nu(1+2\mu)$ is odd (and returns to itself when it is even)
--- and, a feature of the Bannai--Ito grid, never revives fractionally ---
while its isospectral deformation $\alpha\neq\tfrac12$, the rotation of \cite{GTVZ}, converts the
transfer into fractional revival with probabilities $(2\alpha-1)^2$ and $4\alpha(1-\alpha)$ at the two
ends; both are properties of the limiting operator as well, with the same revival time, the mirror surviving as the
sign that labels the two channels $\pm\Qs$. At the algebraic level the two bispectral
operators of the discrete model, the recurrence operator $J$ and the Dunkl--difference operator
$L$, go over in the limit to the differential--difference (Dunkl) realization of the
anticommutator algebra attached to the little $-1$ Jacobi polynomials, with their roles
exchanged: $J$ becomes the Dunkl operator and $L$ the multiplication by the position.

All this is for $N$ odd, i.e.\ an even number $N+1$ of sites. For $N$ even (\S\ref{sec:even}) the matrix is not
chiral --- its diagonal is a constant staggered mass --- and $J$ itself, with no shift,
contracts to the matrix supercharge $\mathcal D_b$ of ordinary supersymmetric quantum mechanics
with the trigonometric P\"oschl--Teller superpotential $\mu\tan x-\tfrac b2\cot x$, whose square is
the shape--invariant pair $\mathcal H_{\mu,b}$, $\mathcal H_{\mu+1,b+2}$, with one unpaired mode that
is exact already at finite $N$. Assembled into one function, this is again a supercharge with reflections of
\cite{PVZ}, with a constant $\mathsf U$ and the P\"oschl--Teller $\mathsf V$, and its square is the Dunkl
P\"oschl--Teller Hamiltonian $\mu(\mu-\R)/\cos^2x+\tfrac b2(\tfrac b2-\R)/\sin^2x$: the two limits
of $J$ are isospectral, the eigenvalues of $J$ being given by the same formulas for both
parities, and the parity of $N$ decides whether the reflection lives in the supercharge or in
the Hamiltonian.

This completes the three continuum limits. The three Hamiltonians realize the same
principle --- a reflection, born of the mirror symmetry that carries state transfer and revival,
combining two channels dictated by the shape of the grid --- at three levels of the Askey scheme from which
the three families are truncated: the
para--Krawtchouk Hamiltonian ($q\to1$, linear grid) gives the isotonic oscillator, with the reflection
selecting a self--adjoint extension; the para--Racah Hamiltonian ($q\to1$, quadratic grid) gives the
trigonometric P\"oschl--Teller Hamiltonian, with the reflection grading two channels; and the
para--Bannai--Ito Hamiltonian ($q\to-1$, two--sided grid) gives the extended Scarf~I Hamiltonian of
supersymmetric quantum mechanics with reflections, the reflection now entering the supercharge itself
as the square root of the Hamiltonian. Each limit is a shape--invariant potential in the sense of
supersymmetric quantum mechanics \cite{CKS}; the para--Bannai--Ito case is the one in which the
supersymmetry is realized with Dunkl operators, as anticipated by the ``square root of the
Schr\"odinger operator'' of the little $-1$ Jacobi polynomials \cite{VZm1} and by the square--root
structures --- the chiral supersymmetry of the $\mathfrak{osp}(1|2)$ Casimir, the Dunkl--Dirac operators ---
that accompany the Bannai--Ito algebra \cite{DGTVZ}.

\section*{Acknowledgements}
NC thanks the Centre de Recherches Math\'ematiques (CRM) for its hospitality. LV is funded in
part through a Discovery Grant of the Natural Sciences and Engineering Research Council (NSERC)
of Canada; SZB, QL, LM and MR enjoy scholarships and fellowships provided by this fund.

\section*{Conflict of interest}
The authors declare that they have no conflict of interest.

\section*{Data availability statement}
No new data were created or analysed in this study.

\appendix
\renewcommand{\theequation}{A.\arabic{equation}}
\setcounter{equation}{0}
\section{The confluence of the para--Bannai--Ito polynomials}
\label{app:confluence}

This appendix supplies the identities used in \S\ref{sec:polylimit} and \S\ref{sec:even}.
Throughout, $\alpha=\half$, $a=-(j+1)-2\mu$, $m_b=\frac{2\mu+b}4$, $m'_b=\mu+\half-m_b$, and
$\sigma=\sin x$ is the position of the site $2m$, $2m=\rho(1+\sigma)$. The asymptotic statements hold
uniformly on compact subsets of $x\in(-\frac\pi2,0)$, i.e.\ away from the walls and from the
centre, and $\simeq$ denotes equality up to a factor $1+O(N^{-1})$.

\subsection{Whipple's transformation.}
\label{app:C1}
The series $F_m$ and $G_m$ of \eqref{eq:F43series} are balanced: the sum of their lower parameters
exceeds the sum of the upper ones by one. For a terminating balanced ${}_4F_3$, Whipple's
transformation \cite[\S7.2]{Bailey} is
\begin{equation}
\label{eq:C:whipple}
{}_4F_3\!\Big(\begin{matrix}-m,\,a_1,\,a_2,\,a_3\\ b_1,\,b_2,\,b_3\end{matrix};1\Big)
=\frac{(b_2-a_1)_m(b_3-a_1)_m}{(b_2)_m(b_3)_m}\;
{}_4F_3\!\Big(\begin{matrix}-m,\,a_1,\,b_1-a_2,\,b_1-a_3\\ b_1,\,a_1+1-m-b_2,\,a_1+1-m-b_3\end{matrix};1\Big),
\end{equation}
valid when $b_1+b_2+b_3=a_1+a_2+a_3-m+1$.
Applied to $F_m$ with $a_1=m-j$, $\{a_2,a_3\}=\{m_b\pm x_s\}$, $b_1=\mu+\half$,
$b_2=\frac{1+b-j}2$ and $b_3=-\frac j2$, it gives, since $b_1-(m_b\pm x_s)=m'_b\mp x_s$,
\begin{equation}
\label{eq:C:FW}
F_m(x_s)=\Pi_m\,F^W_m(x_s),\qquad
F^W_m(x_s)={}_4F_3\!\Big(\begin{matrix}-m,\ m-j,\ m'_b+x_s,\ m'_b-x_s\\ \frac{1-b-j}2,\ \mu+\half,\ 1-\frac j2\end{matrix};1\Big),
\qquad 2m<j,
\end{equation}
the restriction $2m<j$ ensuring that the series terminates before the lower parameter
$1-\frac j2$ produces a zero (it excludes only the central cell, which lies outside every
compact subset of $(-\frac\pi2,0)$), with the prefactor
\begin{equation}
\label{eq:C:Pidef}
\Pi_m=\frac{(\frac{1+b+j}2-m)_m\,(\frac j2-m)_m}{(\frac{1+b-j}2)_m\,(-\frac j2)_m};
\end{equation}
the transformed series $F^W_m$ is $F_m$ with $b\to-b$, $-\frac j2\to1-\frac j2$ and
$m_b\pm x_s\to m'_b\pm x_s=m_{-b}+\tfrac12\pm x_s$.
In the same way $G_m=\Pi'_m\,G^W_m$, where $G^W_m$ is the series $G_m$ with $m'_b\pm x_s$ in place
of $m_b+1\pm x_s$ and with the lower
parameters $\frac{1-b-j}2,\ \mu+\frac32,\ 1-\frac j2$, and $\Pi'_m=\Pi_m\,\frac{ef}{(e+m)(f+m)}$
with $e=\frac{1+b-j}2$, $f=-\frac j2$. Writing the Pochhammer symbols as Gamma functions,
\begin{equation}
\label{eq:C:Pi}
\Pi_m=\frac{\Gamma(\frac{1+b+j}2)\,\Gamma(\frac j2)}{\Gamma(\frac{1-b+j}2)\,\Gamma(\frac j2+1)}\,
\Big(\frac j2-m\Big)\,\frac{\Gamma(\frac{j+1}2-m-\frac b2)}{\Gamma(\frac{j+1}2-m+\frac b2)}
\ \simeq\ c^{W}_N\Big(\frac j2-m\Big)^{1-b}=c^{W}_N\Big(\frac j2\,|\sigma|\Big)^{1-b},
\end{equation}
by $\Gamma(w-\frac b2)/\Gamma(w+\frac b2)\simeq w^{-b}$, the constant $c^{W}_N$ being independent
of $m$ (and distinct from the $c_N$ of \eqref{eq:envelope}); and $\Pi'_m/\Pi_m\simeq 1/\sigma^2$, since $e/(e+m)\simeq f/(f+m)\simeq-1/\sigma$.

\subsection{The coefficients.}
\label{app:C2}
With $(c+1)_m/(c)_m=(c+m)/c$, $(c)_{m+1}/(c)_m=c+m$ and
$(m-j)_m/(m+1-j)_{m+1}=(m-j)/\big((2m-j)(2m+1-j)\big)$, the monic normalizations of
\eqref{eq:F43} give
\begin{equation}
\label{eq:C:kappa}
\begin{aligned}
\frac{\varkappa'_m}{\varkappa_m}&=\frac{(2m+1+b-j)(2m+1+2\mu)(j-m)}{j\,(2\mu+1)(1+b-j)},\\
\frac{\varkappa_{m+1}}{\varkappa_m}&=\frac{(2m+1+b-j)(2m+1+2\mu)(m-j)}{8\,(2m+1-j)},
\end{aligned}
\end{equation}
both exact,
whence $\varkappa'_m/\varkappa_m\simeq-\frac j2\,\frac{\sigma\cos^2x}{2\mu+1}$ and
$\varkappa_{m+1}/\varkappa_m\simeq-\frac{j^2}{16}\cos^2x$. From \eqref{eq:AC}, away from the centre,
$C_{2m}=-\frac m2\,\frac{2m-j-1-b}{2m-j-1}\simeq-\frac j4(1+\sigma)$ and
$C_{2m+1}=-\frac14(2m+1+2\mu)\simeq-\frac j4(1+\sigma)$. Hence
\begin{equation}
\label{eq:C:coef}
-C_{2m}\,\frac{\varkappa'_{m-1}}{\varkappa_m}=-C_{2m}\,\frac{\varkappa'_{m-1}}{\varkappa_{m-1}}\,\frac{\varkappa_{m-1}}{\varkappa_m}
\ \longrightarrow\ \frac{2(1+\sigma)\sigma}{2\mu+1}=c(x),\qquad
C_{2m+1}\,\frac{\varkappa_m}{\varkappa'_m}\ \longrightarrow\ -\frac1{c(-x)},
\end{equation}
the first being \eqref{eq:coeflimit}. The second governs the odd sites: by \eqref{eq:geronimus}
and \eqref{eq:F43},
\begin{equation}
\label{eq:C:odd}
\begin{aligned}
P_{2m+1}(x_s)&=\varkappa'_m\Big[(x_s-m_b)\,G_m(x_s)-C_{2m+1}\frac{\varkappa_m}{\varkappa'_m}\,F_m(x_s)\Big]\\
&\simeq\ -\,\varkappa'_m\,\frac{2\mu+1}{2(1-\sigma)\sigma}\Big[F_m(x_s)+c(-x)\,(x_s-m_b)\,G_m(x_s)\Big]:
\end{aligned}
\end{equation}
the odd--site bracket is the even--site bracket at $-x$. For the model of \S\ref{sec:even}
($N=2j$), where the odd--site amplitude is $\varkappa'_m(x_s-m_b)G_m(x_s)/h_{2m+1}$ with
$h_{2m+1}=h_{2m}u_m$ and $u_m=\sqrt{U_{2m+1}}\simeq\frac j4\cos x$ by \eqref{eq:chaineven},
\begin{equation}
\label{eq:C:evenodd}
\frac{\varkappa'_m}{\varkappa_m\,u_m}\ \longrightarrow\ -\,\frac{2\,\sigma\cos x}{2\mu+1}.
\end{equation}

\subsection{The profile.}
\label{app:C3}
For the model of \S\ref{sec:chain} ($N=2j+1$), \eqref{eq:AC} gives, for $2m+1<j$,
\begin{align*}
U_{2m+1}&=A_{2m}C_{2m+1}=\tfrac1{16}(2j+1+2\mu-2m)(2m+1+2\mu),\\
U_{2m+2}&=A_{2m+1}C_{2m+2}=\frac{(j-m)(m+1)(2m+1-j+b)(2m+1-j-b)}{4\,(2m+1-j)^2},
\end{align*}
so that by \eqref{eq:C:kappa}
\begin{equation}
\label{eq:C:ratio}
\frac{(\varkappa_{m+1}/h_{2m+2})^2}{(\varkappa_m/h_{2m})^2}=\frac{(\varkappa_{m+1}/\varkappa_m)^2}{U_{2m+1}U_{2m+2}}
=\frac{(j-m)\,(2m+1+b-j)\,(2m+1+2\mu)}{(m+1)\,(2m+1-b-j)\,(2j+1+2\mu-2m)}\,,
\end{equation}
and the product from $0$ to $m-1$ is \eqref{eq:envelopeexact}, in agreement with the norm
$h_{2m}$ of \cite[\S4.4]{PBI}. Writing \eqref{eq:envelopeexact} with Gamma functions,
\begin{equation}
\label{eq:C:gamma}
\begin{aligned}
\Big(\frac{\varkappa_m}{h_{2m}}\Big)^2&=c'_N\;\frac{\Gamma(m+\mu+\half)}{\Gamma(m+1)}\;
\frac{\Gamma(j-m+\mu+\frac32)}{\Gamma(j-m+1)}\;
\frac{\Gamma(\frac{j+1+b}2-m)}{\Gamma(\frac{j+1-b}2-m)}\\
&\simeq\ c'_N\;m^{\mu-\frac12}(j-m)^{\mu+\frac12}\Big|\tfrac{j+1}2-m\Big|^{b}\!,
\end{aligned}
\end{equation}
with $c'_N=j!\,\Gamma(\frac{j+1-b}2)\big/\big(\Gamma(\mu+\half)\,\Gamma(j+\mu+\frac32)\,\Gamma(\frac{j+1+b}2)\big)$;
with $m\simeq\frac j2(1+\sigma)$, $j-m\simeq\frac j2(1-\sigma)$ and $\frac{j+1}2-m\simeq\frac j2|\sigma|$,
this is $(1+\sigma)^{\mu-\frac12}(1-\sigma)^{\mu+\frac12}|\sigma|^b=(\cos x)^{2\mu-1}(1-\sigma)\,|\sigma|^b$ up to an
$N$--dependent constant, which is \eqref{eq:envelope}. (Equivalently: in the position
$z=2m+1-j$ of the odd site between the two bonds, the logarithm of \eqref{eq:C:ratio} is
$-\frac{2\mu+1}{j-z}+\frac{2\mu-1}{j+z}+\frac{2b}z+O(N^{-2})$, and consecutive even sites are
two units of $z$ apart.) For the odd sites, \eqref{eq:C:odd} and $h_{2m+1}^2=h_{2m}^2U_{2m+1}$
give the odd--site profile
\begin{equation}
\label{eq:C:oddenv}
\Big[\frac{\varkappa'_m}{h_{2m+1}}\,\frac{2\mu+1}{2(1-\sigma)\sigma}\Big]^2
=\Big(\frac{\varkappa_m}{h_{2m}}\Big)^2\,\frac{(\varkappa'_m/\varkappa_m)^2}{U_{2m+1}}\,\frac{(2\mu+1)^2}{4(1-\sigma)^2\sigma^2}
\ \simeq\ \Big(\frac{\varkappa_m}{h_{2m}}\Big)^2\,\frac{1+\sigma}{1-\sigma},
\end{equation}
by \eqref{eq:C:kappa} and $U_{2m+1}\simeq\frac{j^2}{16}\cos^2x$: the odd--site profile is
$(\cos x)^{\mu-\frac12}|\sigma|^{b/2}(1+\sigma)^{1/2}$, the reflection of the even--site one, with the
\emph{same} constant, as \eqref{eq:channels} requires. For the model of \S\ref{sec:even},
\eqref{eq:C:kappa} and \eqref{eq:chaineven} give $\varkappa_{m+1}/\varkappa_m=-U_{2m+1}$, so that
\eqref{eq:C:ratio} becomes $U_{2m+1}/U_{2m+2}=(u_m/v_m)^2$: the even--site profile is the
unpaired mode \eqref{eq:unpaired} of \S\ref{sec:even}, whose closed form differs from
\eqref{eq:envelopeexact} by $(-j-\half-\mu)_m\to(\half-\mu-j)_m$ in the denominator, i.e.\ by
the factor $(j+\half+\mu)/(j+\half+\mu-m)\simeq\frac{2}{1-\sigma}$, which removes the factor
$(1-\sigma)$: the same computation gives $(\cos x)^{2\mu-1}|\sigma|^b$.

\subsection{The little $-1$ Jacobi identities.}
\label{app:C4}
Two elementary facts are used. First, the connection formula of a terminating Gauss series,
\begin{equation}
\label{eq:C:conn}
{}_2F_1(-n,\mathsf b;\mathsf c;1-\mathsf t)=\frac{(\mathsf c-\mathsf b)_n}{(\mathsf c)_n}\,{}_2F_1(-n,\mathsf b;\mathsf b-\mathsf c-n+1;\mathsf t),
\end{equation}
turns the series in $\cos^2x=1-\sigma^2$ of \eqref{eq:FGlimit} into series in $\mathsf t=\sigma^2$. Write
$F(\mathsf a,\mathsf b;\mathsf c)$ for ${}_2F_1(\mathsf a,\mathsf b;\mathsf c;\mathsf t)$ (sans--serif letters
denote generic hypergeometric parameters) and put
\begin{equation}
\label{eq:C:param}
\mathsf b=n'+2m_b,\quad \mathsf c=\tfrac{b+1}2,\qquad \mathsf b'=n'+2m'_b,\quad \mathsf c'=\tfrac{1-b}2 .
\end{equation}
Then the limits $F,G$ of \eqref{eq:FGlimit} are proportional to $F(-n',\mathsf b;\mathsf c)$ and
$F(1-n',\mathsf b+1;\mathsf c+1)$, the ratio of the two prefactors of \eqref{eq:C:conn} being
$-(2\mu+1)/(b+1)$, while on the second sublattice $F^W,G^W$ are proportional to
$F(-n',\mathsf b';\mathsf c'+1)$ and $F(-n',\mathsf b';\mathsf c')$, with the ratio of prefactors
$(b-1)(2\mu+1)/\big((b-1-2n')(2n'+2\mu+1)\big)$. Second, four contiguous relations,
each a one--line comparison of coefficients:
\begin{equation}
\label{eq:C:contig}
\begin{aligned}
F(\mathsf a,\mathsf b+1;\mathsf c)-F(\mathsf a,\mathsf b;\mathsf c)&=\frac{\mathsf a\,\mathsf t}{\mathsf c}\,F(\mathsf a+1,\mathsf b+1;\mathsf c+1),\\
F(\mathsf a+1,\mathsf b;\mathsf c)-F(\mathsf a,\mathsf b;\mathsf c)&=\frac{\mathsf b\,\mathsf t}{\mathsf c}\,F(\mathsf a+1,\mathsf b+1;\mathsf c+1),\\
\mathsf c\,F(\mathsf a,\mathsf b;\mathsf c)+(\mathsf b-\mathsf c)\,F(\mathsf a,\mathsf b;\mathsf c+1)&=\mathsf b\,F(\mathsf a,\mathsf b+1;\mathsf c+1),\\
\mathsf c\,F(\mathsf a,\mathsf b;\mathsf c)+(\mathsf a-\mathsf c)\,F(\mathsf a,\mathsf b;\mathsf c+1)&=\mathsf a\,F(\mathsf a+1,\mathsf b;\mathsf c+1).
\end{aligned}
\end{equation}
\emph{First sublattice.} By the above, the bracket of \eqref{eq:limitform} is proportional to
\[
F(-n',\mathsf b;\mathsf c)-\frac{(1+\sigma)\,\sigma\,(x_s-m_b)}{\mathsf c}\,F(1-n',\mathsf b+1;\mathsf c+1).
\]
For
$x_s=m_b+n'$, i.e.\ $x_s-m_b=n'$, the first relation of \eqref{eq:C:contig} with $\mathsf a=-n'$
absorbs the $\sigma^2$ term and leaves
$F(-n',\mathsf b+1;\mathsf c)-\frac{2n'\sigma}{b+1}\,F(1-n',\mathsf b+1;\mathsf c+1)$, which is
$P^{(b,2\mu)}_{2n'}(-\sigma)$ by \eqref{eq:m1explicit}, since $\mathsf b+1=n'+\frac{b+2\mu}2+1$. For
$x_s=-(m_b+n')$, i.e.\ $x_s-m_b=-\mathsf b$, the second relation absorbs it and leaves
$F(1-n',\mathsf b;\mathsf c)+\frac{2\mathsf b \sigma}{b+1}\,F(1-n',\mathsf b+1;\mathsf c+1)$, which is
$P^{(b,2\mu)}_{2n'-1}(-\sigma)$, since $2\mathsf b=b+2\mu+2n'$. This is \eqref{eq:bracket}.
\emph{Second sublattice.} For $x_s=m'_b+n'$ one has $x_s-m_b=n'+\mathsf c'$, and
$\sigma F^W+\frac{2(1+\sigma)}{2\mu+1}(x_s-m_b)G^W$ is proportional to
\[
F(-n',\mathsf b';\mathsf c')+\sigma\Big[F(-n',\mathsf b';\mathsf c')+\frac{\mathsf b'-\mathsf c'}{\mathsf c'}F(-n',\mathsf b';\mathsf c'+1)\Big];
\]
the third relation turns the square bracket into $\frac{\mathsf b'}{\mathsf c'}F(-n',\mathsf b'+1;\mathsf c'+1)$,
and $F(-n',\mathsf b';\mathsf c')+\frac{2\mathsf b'\sigma}{1-b}F(-n',\mathsf b'+1;\mathsf c'+1)$ is
$P^{(-b,2\mu)}_{2n'+1}(-\sigma)$, since $2\mathsf b'=2n'+2+2\mu-b$. For $x_s=-(m'_b+n')$ one has
$x_s-m_b=-(n'+\mu+\half)$, the same combination is proportional to
\[
F(-n',\mathsf b';\mathsf c')+\sigma\Big[F(-n',\mathsf b';\mathsf c')-\frac{n'+\mathsf c'}{\mathsf c'}F(-n',\mathsf b';\mathsf c'+1)\Big],
\]
the fourth relation turns the square bracket into $-\frac{n'}{\mathsf c'}F(1-n',\mathsf b';\mathsf c'+1)$,
and $F(-n',\mathsf b';\mathsf c')-\frac{2n'\sigma}{1-b}F(1-n',\mathsf b';\mathsf c'+1)$ is $P^{(-b,2\mu)}_{2n'}(-\sigma)$.
This is \eqref{eq:bracketW}.


\begin{thebibliography}{99}

\bibitem{VZ} L.\ Vinet and A.\ Zhedanov, \emph{Para--Krawtchouk polynomials on a bi--lattice
and a quantum spin chain with perfect state transfer}, J.\ Phys.\ A: Math.\ Theor.\
\textbf{45}, 265304 (2012).

\bibitem{PKcont} S.\ Z.\ Beaulac, N.\ Cramp\'e, Q.\ Labriet, L.\ Morey, M.\,J.\ Rivera Valladares and L.\ Vinet, \emph{The
para--Krawtchouk oscillator and its continuum limit} (2026), companion paper, in preparation.

\bibitem{PRcont} S.\ Z.\ Beaulac, N.\ Cramp\'e, Q.\ Labriet, L.\ Morey, M.\,J.\ Rivera Valladares and L.\ Vinet, \emph{The para--Racah Hamiltonian as an exactly solvable discretization of the P\"oschl--Teller box}
(2026), companion paper, in preparation.

\bibitem{paraRacah} J.-M.\ Lemay, L.\ Vinet and A.\ Zhedanov, \emph{The para--Racah
polynomials}, J.\ Math.\ Anal.\ Appl.\ \textbf{438}, 565 (2016).

\bibitem{paraRacahChain} J.-M.\ Lemay, L.\ Vinet and A.\ Zhedanov, \emph{An analytic spin
chain model with fractional revival}, J.\ Phys.\ A: Math.\ Theor.\ \textbf{49}, 335302 (2016).

\bibitem{GVZ} V.\,X.\ Genest, L.\ Vinet and A.\ Zhedanov, \emph{Quantum spin chains with
fractional revival}, Ann.\ Phys.\ \textbf{371}, 348 (2016).

\bibitem{AP} I.\,Sh.\ Averbukh and N.\,F.\ Perelman, \emph{Fractional revivals: universality in the
long--term evolution of quantum wave packets beyond the correspondence principle dynamics}, Phys.\
Lett.\ A \textbf{139}, 449 (1989).

\bibitem{Robinett} R.\,W.\ Robinett, \emph{Quantum wave packet revivals}, Phys.\ Rep.\ \textbf{392},
1 (2004).

\bibitem{PBI} J.\ Pelletier, L.\ Vinet and A.\ Zhedanov, \emph{Para--Bannai--Ito polynomials},
SIGMA \textbf{19}, 090 (2023).

\bibitem{BCLMNV} S.\ Z.\ Beaulac, N.\ Cramp\'e, Q.\ Labriet, L.\ Morey, R.\,I.\ Nepomechie and
L.\ Vinet, \emph{Para families of orthogonal polynomials} (2026), in preparation.

\bibitem{VZm1} L.\ Vinet and A.\ Zhedanov, \emph{A ``missing'' family of classical orthogonal
polynomials}, J.\ Phys.\ A: Math.\ Theor.\ \textbf{44}, 085201 (2011).

\bibitem{PVZ} S.\ Post, L.\ Vinet and A.\ Zhedanov, \emph{Supersymmetric quantum mechanics
with reflections}, J.\ Phys.\ A: Math.\ Theor.\ \textbf{44}, 435301 (2011).

\bibitem{DGTVZ} H.\ De Bie, V.\,X.\ Genest, S.\ Tsujimoto, L.\ Vinet and A.\ Zhedanov, \emph{The
Bannai--Ito algebra and some applications}, J.\ Phys.: Conf.\ Ser.\ \textbf{597}, 012001 (2015).

\bibitem{TVZ} S.\ Tsujimoto, L.\ Vinet and A.\ Zhedanov, \emph{Dunkl shift operators and
Bannai--Ito polynomials}, Adv.\ Math.\ \textbf{229}, 2123 (2012).

\bibitem{GTVZ} V.\,X.\ Genest, S.\ Tsujimoto, L.\ Vinet and A.\ Zhedanov, \emph{Persymmetric
Jacobi matrices, isospectral deformations and orthogonal polynomials}, J.\ Math.\ Anal.\ Appl.\
\textbf{450}, 915 (2017).

\bibitem{BI} E.\ Bannai and T.\ Ito, \emph{Algebraic Combinatorics I: Association Schemes},
Benjamin/Cummings, Menlo Park (1984).

\bibitem{Dunkl} C.\,F.\ Dunkl, \emph{Differential--difference operators associated to reflection
groups}, Trans.\ Amer.\ Math.\ Soc.\ \textbf{311}, 167 (1989).

\bibitem{CKS} F.\ Cooper, A.\ Khare and U.\ Sukhatme, \emph{Supersymmetry and quantum
mechanics}, Phys.\ Rep.\ \textbf{251}, 267 (1995).

\bibitem{KLS} R.\ Koekoek, P.\,A.\ Lesky and R.\,F.\ Swarttouw, \emph{Hypergeometric
Orthogonal Polynomials and Their $q$--Analogues}, Springer (2010).

\bibitem{Bailey} W.\,N.\ Bailey, \emph{Generalized Hypergeometric Series}, Cambridge Tracts in
Mathematics and Mathematical Physics \textbf{32}, Cambridge University Press, Cambridge, 1935.

\bibitem{SSH} W.\,P.\ Su, J.\,R.\ Schrieffer and A.\,J.\ Heeger, \emph{Solitons in
polyacetylene}, Phys.\ Rev.\ Lett.\ \textbf{42}, 1698 (1979).

\bibitem{TLM} H.\ Takayama, Y.\,R.\ Lin--Liu and K.\ Maki, \emph{Continuum model for solitons
in polyacetylene}, Phys.\ Rev.\ B \textbf{21}, 2388 (1980).

\bibitem{JR} R.\ Jackiw and C.\ Rebbi, \emph{Solitons with fermion number $\tfrac12$}, Phys.\ Rev.\ D
\textbf{13}, 3398 (1976).

\bibitem{CLMPV} N.\ Cramp\'e, Q.\ Labriet, L.\ Morey, G.\ Parez and L.\ Vinet,
\emph{Inhomogeneous SSH models and the doubling of orthogonal polynomials},
SciPost Phys.\ \textbf{20}, 124 (2026), arXiv:2511.15527.

\bibitem{PCLMV} G.\ Parez, N.\ Cramp\'e, Q.\ Labriet, L.\ Morey and L.\ Vinet,
\emph{Universal scaling of spatially extended zero modes in inhomogeneous SSH chains},
arXiv:2608.11021.

\bibitem{OVdJ1} R.\ Oste and J.\ Van der Jeugt, \emph{Doubling (dual) Hahn polynomials:
classification and applications}, SIGMA \textbf{12}, 003 (2016).

\bibitem{Rosenblum} M.\ Rosenblum, \emph{Generalized Hermite polynomials and the Bose--like
oscillator calculus}, Oper.\ Theory Adv.\ Appl.\ \textbf{73}, 369 (1994).

\bibitem{JNT} W.\,B.\ Jones, O.\ Nj\aa stad and W.\,J.\ Thron, \emph{Moment theory, orthogonal
polynomials, quadrature, and continued fractions associated with the unit circle}, Bull.\
London Math.\ Soc.\ \textbf{21}, 113 (1989).

\end{thebibliography}
\end{document}